\documentclass[11pt]{article} 
\pdfoutput=1 

\usepackage{jheppub} 

\usepackage[utf8]{inputenc}
\usepackage[english]{babel}
\usepackage{amsmath,amssymb,amsbsy,amstext,amsthm,booktabs,simplewick,exscale,relsize,slashed,graphicx,amsfonts,upgreek, xcolor,subcaption}
\usepackage{multirow,color,dcolumn,bm,enumerate}
\usepackage{hyperref}
\usepackage{feynmp-auto,axodraw2,tikz}
\usepackage{ulem}
\usepackage{comment}
\usepackage{amsmath}
\usepackage{ytableau}
\DeclareUnicodeCharacter{2212}{-}

\usepackage{hyperref}
\usepackage[capitalise,noabbrev,nameinlink]{cleveref}

\title{A case study of SMEFT $\mathcal O(1/\Lambda^4)$ effects in diboson processes: $pp \to W^\pm(\ell^\pm \nu) \gamma$}
\author[a]{Adam Martin,}

\affiliation[a]{Department of Physics, University of Notre Dame,
  South Bend, IN, 46556 USA}

\emailAdd{amarti41@nd.edu}

\abstract{
In this paper we explore $pp \to W^\pm (\ell^\pm \nu) \gamma$ to $\mathcal O(1/\Lambda^4)$ in the SMEFT expansion. Calculations to this order are necessary to properly capture SMEFT contributions that grow with energy, as the interference between energy-enhanced SMEFT effects at $\mathcal O(1/\Lambda^2)$  and the Standard Model is suppressed. We find that there are several dimension eight operators that interfere with the Standard Model and lead to the same energy growth, $\sim \mathcal O(E^4/\Lambda^4)$, as dimension six squared. While energy-enhanced SMEFT contributions are a main focus, our calculation includes the complete set of $\mathcal O(1/\Lambda^4)$ SMEFT effects consistent with $U(3)^5$ flavor symmetry. Additionally, we include the decay of the $W^\pm \to \ell^\pm\nu$, making the calculation actually $\bar q q' \to \ell^\pm \nu \gamma$. As such, we are able to study the impact of non-resonant SMEFT operators, such as $(L^\dag\bar\sigma^\mu \tau^I\, L)  (Q^\dag\bar\sigma^\nu \tau^I\, Q)\, B_{\mu\nu}$, which contribute to $\bar q q' \to \ell^\pm \nu \gamma$ directly and not to $\bar q q' \to W^\pm \gamma$. We show several distributions to illustrate the shape differences of the different contributions. 
}

\begin{document}
\maketitle

\setcounter{page}{2}

\section{Introduction}\label{sec:intro}

In this paper we explore $pp \to W^\pm \gamma$ production to $\mathcal O(1/\Lambda^4)$ with the Standard Model Effective field theory expansion (SMEFT)~\cite{Buchmuller:1985jz, Grzadkowski:2010es,Brivio:2017vri}. The SMEFT framework extends the SM by a series of higher dimensional operators (formed from the SM fields and their derivatives),
\begin{align}
\mathcal{L}_{S M E F T}=\mathcal{L}_{S M}+\sum_{d=5} \sum_{i=1}^n \frac{C_i^{(d)} \mathcal{O}_i^{(d)}\left(Q, u_c, d_c, L, e_c, H, B_{\mu \nu}, W_{\mu \nu}^I, G_{\mu \nu}^A ; D_\mu\right)}{\Lambda^{d-4}}
\label{eq:SMEFT}
\end{align}
where $d$ labels the operator mass dimension, and $i$ enumerates the independent operators at a given mass dimension\footnote{While the sum in Eq.~\eqref{eq:SMEFT} begins at $d =5$, in this work  will assume that baryon number and lepton number are preserved. This removes odd dimensions from the sum, so that $d=6$ is the lowest dimension.}. The scale $\Lambda$ is assumed to be $\gg$ all other masses and energies; we take it to have the same value for all operators, encoding any relative differences in the strengths operators into the Wilson coefficients $C^{(d)}_i$. While any UV scenario with a single source of electroweak symmetry breaking and with all states far heavier than the SM can be mapped into the SMEFT, our interest will be to use it from the bottom up, keeping all operators to a given order in $1/\Lambda$ and with minimal assumptions about the relative sizes of the $C^{(d)}_i$.

 Diboson processes such as $pp \to W\gamma$ are key LHC process from the SMEFT perspective as they provide insight on the nature of the triple gauge couplings $VVV$. The first constraints on the $VVV$ couplings come from LEPII~\cite{ALEPH:2004dmh, L3:2004lwm, OPAL:2007ytu, ALEPH:2013dgf}, with updates from studies at the Tevatron and LHC, e.g. Ref~\cite{CMS:2020mxy, ATLAS:2019rob, ATLAS:2019bsc, CMS:2021icx, ATLAS:2013way, CMS:2020olm}. Phrased in the language of operators/Wilson coefficients, fermion pair production, either from $e^+e^- \to \bar f f$ at LEP or Drell-Yan at the Tevatron/LHC set strong constraints on SMEFT operators that affect fermion-fermion-gauge boson couplings $ffV$, but have no impact on the operator $C^{(6)}_W\, \epsilon_{IJK} W_\mu^{I,\nu}\,W_\nu^{J\rho}W_\rho^{K\mu}$  and only indirect (via input parameter dependence) on $C^{(6)}_{HWB} H^\dag \tau^I H\, W^I_{\mu\nu}B^{\mu\nu}$ (working in the Warsaw~\cite{Grzadkowski:2010es} basis, and considering tree-level effects only).
 
 Given its role in constraining new SMEFT operators and thereby helping to piece together a global picture of deviations from the SM, it is important to provide accurate theoretical predictions. Usual perturbative SM corrections fall in this category, but the purpose of this paper is to study corrections from higher order terms in the SMEFT expansion. The lowest order terms come from dimension six operators interfering with the SM, $\mathcal O(1/\Lambda^2)$, and have been studied extensively; see Ref~\cite{Butter:2016cvz, Green:2016trm, Zhang:2016zsp, Baglio:2017bfe,Ellis:2018gqa,Baglio:2018bkm,Baglio:2019uty,Bellan:2021dcy,Almeida:2021asy} for diboson-focused analyses. The next highest term in the SMEFT expansion is $\mathcal O(1/\Lambda^4)$, which can either come from the product of two dimension six contributions (either the square of a single operator, which we'll call `self-square' terms, or the product of two different operators), or from the interference between a dimension eight operator and the SM. 
 
 One motivation to look at higher orders comes from energetics. When SMEFT operators are added to an amplitude, they turn into contributions of $\mathcal O(E^{d-4}/\Lambda^{d-4})$, where $E$ is roughly the partonic center of mass energy of the process\footnote{For operators containing Higgses, the energy growth can be slower, with powers of the Higgs vev replacing powers of energy. In those cases,  $\mathcal O(E^{d-4}/\Lambda^{d-4})$ is the leading energy behavior.}. As such, a natural place to look for SMEFT effects is at high energy, in the tails of kinematic distributions. However, the higher the energy, the larger the expansion parameter $E/\Lambda$ and the larger the impact from higher order terms.

For diboson processes, there is reason to believe the $\mathcal O(1/\Lambda^4)$ effects are particularly important. The operator that leads to the strongest energy growth at dimension six is $C^{(6)}_W\, \epsilon_{IJK} W_\mu^{I,\nu}\,W_\nu^{J\rho}W_\rho^{K\mu}$\footnote{In this paper, we will often use the shorthand of referring to operators by their Wilson coefficients, e.g. $C^{(6)}_W\, \epsilon_{IJK} W_\mu^{I,\nu}\,W_\nu^{J\rho}W_\rho^{K\mu}$ becomes $C^{(6)}_W$. Our convention for operators is explained in Sec.~\ref{sec:SMEFTcont}. }, with $A_{C_W} \sim E^2/\Lambda^2$. However, the enhanced part of $A_{C_W}$ involves different $W/\gamma$ polarizations than the dominant SM piece (in the limit $E \gg v$), so the interference between the two is suppressed~\cite{Azatov:2016sqh}\footnote{The suppressed interference holds at the level of the total cross section. Interference can be revived to some extent by looking at the azimuthal distributions of the reconstructed vector bosons~\cite{Azatov:2017kzw,Panico:2017frx} }. To see the energy enhancement from $C^{(6)}_W$, one needs to work to $\mathcal O(1/\Lambda^4)$ and include the self square, $|A_{C_W}|^2 \sim E^4/\Lambda^4$. However, once we look to higher order, we must consider all $\mathcal O(1/\Lambda^4)$ effects, and, in particular, dimension eight operators that lead to energy enhanced amplitudes in the polarization channels where the SM is largest. A set of these dimension eight operators and their effect on $WW$ and $WZ$ has been discussed in Ref.~\cite{Degrande:2023iob}, motivated by this large $E$ argument.

In addition to studying a different diboson process, our calculation includes the full $\mathcal O(1/\Lambda^4)$ calculation consistent with $U(3)^5$ flavor and CP symmetry. Our motivation for the full calculation is that, while energy arguments are a good starting point, they only involve a subset of the operators and may be offset by hierarchies among coefficients. We also decay the $W^\pm$, so that the full SMEFT calculation is $pp \to \ell^\pm \nu \gamma$. Compared to $pp \to W^\pm\gamma$, the $2 \to 3$ calculation includes non-resonant effects, meaning operators that contribute to $\ell^\pm \nu \gamma$ without an intermediate, on-shell $W$. From energy arguments, several of these non-resonant effects, such as a $\bar q q' \ell^\pm \nu$ contact interaction with a photon emitted off the $\bar q, q'$ or $\ell^\pm$ lines, are energy enhanced and could complicate interpretations of high energy $\ell^\pm \nu \gamma$ events in terms of triple gauge vertices. Prior studies~\cite{Corbett:2023yhk} have assumed that the non-resonant contributions can be controlled with analysis cuts, but this has not been explicitly checked.

There are several advantages to using $W^\pm(\ell^\pm \nu)\gamma$ as a laboratory for $\mathcal O(1/\Lambda^4)$ effects. First, only left handed fermions participate in the SM amplitude, and since dimension eight operators must interfere with the SM to contribute at $\mathcal O(1/\Lambda^4)$, we only need to consider dimension eight operators containing left handed fermions. Second, the final state is charged, so there are no gluon initiated effects (at the $2 \to 3$ level). Finally, the final state including $W^\pm$ decays is three-body, while cousin processes like $W^+W^-$ or $W^\pm Z$ are four-body when fully decayed. 

The rest of this paper is structured as follows. In Sec.~\ref{sec:SMEFTcont} we break down the various tree level contributions to $pp \to W^\pm(\ell^\pm\nu)\gamma$ into subcategories depending on the topology of the Feynman graphs and whether or not the $\ell^\pm\nu$ system is resonant. For each subcategory, we derive the helicity amplitudes (taking all fermions to be massless) in Sec.~\ref{sec:theamps}, followed by the corresponding coupling factors in Sec.~\ref{sec:coupling}, noting the order in the SMEFT expansion where each appears. These amplitudes and their accompanying coupling factors are the main result of this paper. Before turning to numerics, in Sec.~\ref{sec:pol} we study the polarization breakdown of the resonant terms, as this controls which SMEFT effects are dominant in the high-energy region ($\sqrt{\hat s} \gg v$). We then explore the ($\sqrt{\hat s} \gg v$) regime numerically in Sec.~\ref{sec:results}, showing how non-resonant terms can be mitigated by additional cuts. This section also contains some kinematic distributions, which further illustrate the similarities and differences between different SMEFT operator effects. Section~\ref{sec:conclude} contains our conclusions.

\section{SMEFT contributions to $pp \to W^\pm(\ell^\pm) \gamma$}\label{sec:SMEFTcont}

SMEFT effects enter into $pp \to W^\pm(\ell^\pm) \gamma$ in three ways:
\begin{itemize}
\item Through input parameter dependence, the way SM inputs such as $g_1, g_2, v$ are connected to experimental data.
\item Through altered three-particle vertices, specifically those involving two fermions and a vector boson ($\bar q qW^\pm, \ell^\pm \nu W^\pm$ and $ff \gamma, f = q, \ell^\pm$) or three vector bosons $VVV = W^+W^-\gamma$. The $ff\gamma$ vertex remains unchanged (when the fermions have the same chirality) in SMEFT due to gauge invariance. All other three-particle vertices receive SMEFT corrections.
\item Through contact four and five-particle vertices $\bar q q' W^\pm \gamma, \ell^\pm \nu W^\pm \gamma, \bar q q' \ell^\pm \nu$ and $\bar q q' \ell^\pm \nu \gamma$. These have no SM analog and are purely SMEFT effects. Note that the five particle vertices lead to `non-resonant' contributions.
\end{itemize}
Within the geoSMEFT framework, the first two types of SMEFT effects have already conveniently been set -- meaning the set of possible kinematic forms and the operators that can contribute have been enumerated and grouped into field-dependent `metrics'. As explained in Ref.~\cite{Helset:2020yio}, these contributions are functions of $v/\Lambda$ alone and can therefore be expressed in compact, all-orders forms. To get the $\mathcal O(1/\Lambda^4)$ contributions, we simply need to expand the metrics to that order.  What remains is to find all the four and five-particle interactions that can affect  $\bar q q' \to W^\pm(\ell^\pm) \gamma$. A few four and five-particle interactions are contained within the geoSMEFT `metrics' \footnote{For example, the operator $i(Q^\dag \bar \sigma^\mu Q)(H^\dag \overleftrightarrow{D}_\mu H)$ contains a correction to the $\bar q q Z$ vertex if both Higgses are set to their vevs, hence it is a member of the geoSMEFT metric $L^\psi_{I,A}$, but it also generates $\bar q q Z h, \bar q q Z h^2$, etc. vertices}, however to capture all four and five-particle effects we will need to look beyond geoSMEFT.

We will first sort SMEFT effects diagrammatically, then connect to operators. The set of diagrams are presented below in Figs~\ref{fig:resonant} and \ref{fig:nonresonant}, grouped into whether the $\ell^\pm \nu$ system is resonant or not.

\begin{figure*}[h!]
\centering
\includegraphics[width=0.75\textwidth]{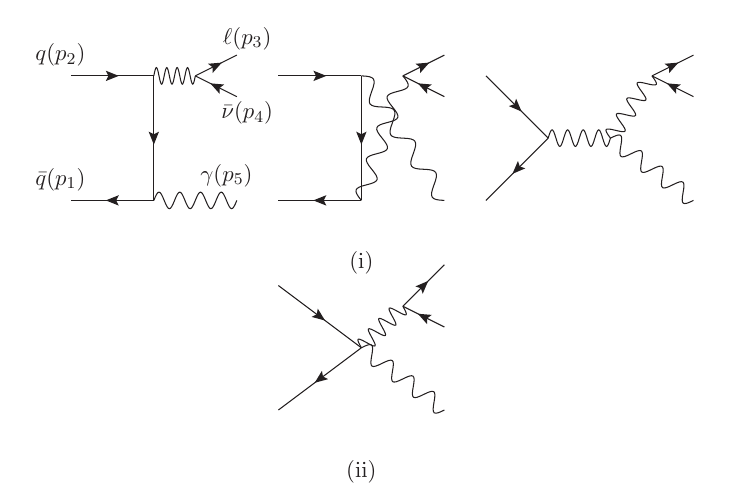}
\caption{Tree level resonant topologies for $pp \to W^-(\ell^- \bar\nu)\gamma$ in SMEFT. External labels for all diagrams are the same as in the upper left diagram, and all momentum have been taken to be outgoing. }
\label{fig:resonant}
\end{figure*}

\begin{figure*}[h!]
\centering
\includegraphics[width=0.75\textwidth]{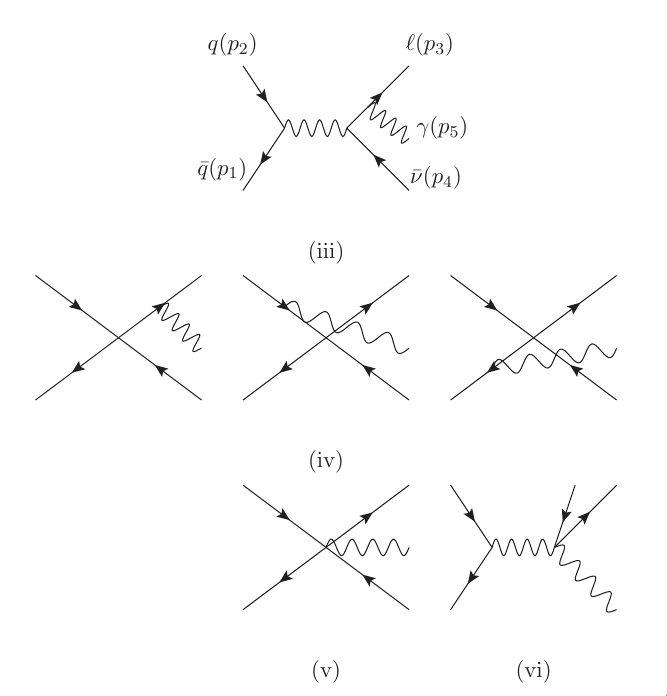}
\caption{Tree level nonresonant topologies for $pp \to W^-(\ell^- \bar \nu)\gamma$ in SMEFT. External labels for all diagrams are the same as in the labeled diagram, and all momentum have been taken to be outgoing. }
\label{fig:nonresonant}
\end{figure*}

Diagrams i.) and iii.) show the SM contributions. Extended to SMEFT, the couplings that accompany these diagrams will contain effects from higher dimensional operators. In the terminology of Ref.~\cite{Helset:2020yio} the metrics appearing are the $h_{IJ}(\phi)$, $g_{AB}(\phi)$ and $L^\psi_{I,A}(\phi)$, which correspond to corrections to the Higgs kinetic term, the $W/Z/\gamma$ kinetic term, and the coupling of fermionic currents to gauge bosons, respectively. Here, $\phi^2 \equiv H^\dag H/\Lambda^2$, the indices $I,J$ label the four degrees of freedom in the Higgs, $A,B$ label the four gauge bosons, and $\psi$ indicates the fermion type. See Ref.~\cite{Helset:2020yio} for the expansions of these metrics in terms of higher dimensional operators, or Ref.~\cite{Kim:2022amu} for the coupling shifts as a function of the Wilson coefficients up to $\mathcal O(1/\Lambda^4)$. For the triboson coupling $VVV$, SMEFT effects will also introduce new kinematics. Despite their non-SM kinematics, these interactions still lie in the geoSMEFT framework, meaning they can be expressed as a Higgs-dependent metric multiplying three field strengths $f_{IJK}(\phi) W_{\mu}^{I,\nu}W^{J,\rho}_\nu W^{K,\mu}_\rho$, with the form of $f_{IJK}(\phi)$ known to all orders in $\phi$. Note that no derivatives appear in any of the geoSMEFT metrics (they are e.g. $h_{IJ}(\phi)$, not $h_{IJ}(\phi, \partial)$), as terms with derivatives either reduce to existing terms or only contribute to vertices with more than three particles~\cite{Helset:2020yio}.

Diagram ii.)  contains a  $\bar q q' W^\pm \gamma$ contact interaction. This does not occur at dimension six, at least within the Warsaw basis. At dimension eight, operators in the class\footnote{By class we are referring to the field and derivative content only. When discussing operator classes, we will follow the convention of Red~\cite{Grzadkowski:2010es,Murphy:2020rsh}, using $X$ to denote any field strength and $D$ for powers of derivatives. We use $Q,L, u_c$, etc. to indicate the fermion type, but won't bother distinguishing fermions from antifermions.} $Q^2 X^2 D$ and $Q^2 X H^2 D$ will contribute; $Q^2 X^2 D$ clearly contains the field content required, while for $Q^2 X H^2 D$, we get a contribution when the derivative is placed on the Higgs field and the Higgs is set to its vev, i.e. $D_\mu H \supset g\, v\, W^\pm_\mu$. Note we do not consider operators with right handed fermions, i.e. $D u_c^2 H^2 X$ since dimension eight operators must interfere with the SM amplitude to contribute to the cross section at $\mathcal O(1/\Lambda^4)$ and the SM amplitude only involves left handed fermions. One may have expected operator class $Q^2H^2 D^3$ to contribute, however once these operators are expanded out we find no $\bar q q' W^\pm \gamma$ vertices. This is not surprising given that $Q^2H^2D^3$ has no field strength $X$ to represent a transverse photon. Had we explored $pp \to W^\pm Z$,  $Q^2H^2D^3$ would contribute to the longitudinal $W/Z$ polarizations (see Ref~\cite{Degrande:2023iob}). Diagram vi.) is the leptonic version of diagram ii.) and is affected by $L^2 X^2 D$ and $L^2 X H^2 D$ operators.  

The remaining two sets of diagrams ((iii.), (iv.)) contain four fermion vertices at their core. Diagram iv.) is sensitive to any four fermion operator containing the right field content at dimension six, while at dimension eight only four fermion operators with left handed fermions are relevant, $Q^2 L^2 H^2 $ and $Q^2 L^2 D^2 $. The fact that the chirality (and flavor) of the dimension six operators entering diagram iv.) is not affected by the requirement of interference with the SM leads to a proliferation of potential operators. To tame this abundance of operators and focus on questions of energy enhanced amplitudes and the role of non-resonant contributions, we will impose $U(3)^5$ flavor symmetry as defined in Ref.~\cite{Brivio:2017btx}. This assumption, eliminates all four fermion operators except those with $Q^\dag Q L^\dag L$ form (only left handed fields). It also removes dipole operators and right handed $W^\pm$ currents, both of which would otherwise contribute to diagrams i.) and iii.). Breaking the $U(3)^5$ by reinstating third-generation Yukawa couplings, or by moving to full minimal flavor violation~\cite{DAmbrosio:2002vsn}, does not admit any additional operators.


Finally, diagram v.) contains the five-particle contact terms. These can be generated by the $Q^2 L^2 D^2 $ operators via the covariant derivative, or by operators in the $Q^2L^2X$ class.

Based off of this enumeration, the list of operators that contribute to the four and five-particle diagrams are listed below in Table~\ref{fig:theoperators}. The operators contributing to three-particle vertices. and input parameter dependence have already been listed in Ref.~\cite{Helset:2020yio, Kim:2022amu}. We impose $U(3)^5$ flavor symmetry, as explained, as well as CP symmetry, given that CP violating interactions are tightly constrained by low-energy experiments~\cite{Cirigliano_2016}. The operators are listed by mass dimension and class/field content.

\begin{table}[h!]
\centering
\addtolength{\leftskip} {-2cm}
\addtolength{\rightskip}{-2cm}
\begin{tabular}{|cl|cl|} \hline
\multicolumn{2}{|c|}{Class $\psi^2XH^2D$, $\psi = Q,L$} & \multicolumn{2}{|c|}{Class $\psi^2X^2D$, $\psi = Q,L$} \\ \hline
\multicolumn{2}{|c|}{}  & \multicolumn{2}{|c|}{}  \\
$C^{(8),1}_{\psi^2BH^2D}$ & $(\psi^\dag \bar\sigma^\nu \tau^I \psi)D^\mu(H^\dag \tau^I\, H)\, B_{\mu\nu}$ & $C^{(8),2}_{\psi^2W^2D}$ & $\epsilon_{IJK}(\psi^\dag \bar\sigma^\mu \overleftrightarrow{D}^\nu \tau^I \psi)\, W^J_{\mu\rho}W^K_{\nu\rho}$ \\
$C^{(8),3}_{\psi^2BH^2D}$ & $i\,(\psi^\dag \bar\sigma^\nu \tau^I \psi)(H^\dag \overleftrightarrow{D}^\mu \tau^I\, H)\, B_{\mu\nu}$ & $C^{(8),1}_{\psi^2WBD}$ & $i\,(\psi^\dag \bar\sigma^\mu \overleftrightarrow{D}^\nu \tau^I \psi)\, (B_{\mu\rho}W^{I,\rho}{}_{\nu} - B_{\nu\rho}W^{I,\rho}{}_{\mu})$\\
$C^{(8),9}_{\psi^2WH^2D}$ & $\epsilon_{IJK}(\psi^\dag \bar\sigma^\nu \tau^I \psi)\, D^\mu(H^\dag \tau^J\, H)\, W^K_{\mu\nu}$ & $C^{(8),2}_{\psi^2WBD}$ & $i\,(\psi^\dag \bar\sigma^\mu \overleftrightarrow{D}^\nu \tau^I \psi)\, (B_{\mu\rho}W^{I,\rho}{}_{\nu} + B_{\nu\rho}W^{I,\rho}{}_{\mu})$ \\
$C^{(8),11}_{\psi^2WH^2D}$ &  $i\, \epsilon_{IJK}(\psi^\dag \bar\sigma^\nu \tau^I \psi)\, (H^\dag \overleftrightarrow{D}^\mu \tau^J\, H)\, W^K_{\mu\nu}$ & \multicolumn{2}{|c|}{} \\
\multicolumn{2}{|c|}{}  & \multicolumn{2}{|c|}{}   \\ \hline
\multicolumn{2}{|c|}{Class $Q^2L^2$} & \multicolumn{2}{|c|}{Class $Q^2L^2D^2$} \\ \hline 
\multicolumn{2}{|c|}{}  & \multicolumn{2}{|c|}{} \\
$C^{(6),3}_{Q^2L^2}$ & $(Q^\dag\bar\sigma^\mu \tau^I\, Q)(L^\dag\bar\sigma_\mu \tau^I\, L)$ & $C^{(8),s,3}_{Q^2L^2}$ & $D^2(Q^\dag\bar\sigma^\mu \tau^I\, Q)(L^\dag\bar\sigma_\mu \tau^I\, L)$ \\ 
\multicolumn{2}{|c|}{}  & $C^{(8),t,3}_{Q^2L^2}$ & $(D^\nu Q^\dag\bar\sigma^\mu \tau^I\, Q)(D_\nu L^\dag\bar\sigma_\mu \tau^I\, L) + h.c.$  \\
\multicolumn{2}{|c|}{}  & \multicolumn{2}{|c|}{}   \\ \hline
\multicolumn{2}{|c|}{Class $Q^2L^2H^2$} & \multicolumn{2}{|c|}{Class $Q^2L^2X$} \\ \hline
\multicolumn{2}{|c|}{}  & \multicolumn{2}{|c|}{}   \\ 
$C^{(8),2}_{Q^2L^2H^2}$ & $(H^\dag H)(Q^\dag\bar\sigma^\mu \tau^I\, Q)(L^\dag\bar\sigma_\mu \tau^I\, L)$ & $C^{(8),5}_{L^2Q^2W}$ & $\epsilon_{IJK}  (L^\dag\bar\sigma^\mu \tau^I\, L)  (Q^\dag\bar\sigma^\nu \tau^J\, Q)\, W^K_{\mu\nu}$  \\
$C^{(8),5}_{Q^2L^2H^2}$ & $\epsilon_{IJK}(H^\dag\, \tau^I H)(Q^\dag\bar\sigma^\mu \tau^J\, Q)(L^\dag\bar\sigma_\mu \tau^K\, L)$ & $C^{(8),3}_{L^2Q^2B}$ & $(L^\dag\bar\sigma^\mu \tau^I\, L)  (Q^\dag\bar\sigma^\nu \tau^I\, Q)\, B_{\mu\nu}$ \\
\multicolumn{2}{|c|}{}  & \multicolumn{2}{|c|}{}   \\ \hline
\end{tabular}
\caption{Operators at dimension six and eight that contribute to contact terms in the process $pp \to W^\pm (\ell^\pm\nu)\gamma$. We have imposed CP conservation and $U(3)^5$ flavor symmetry to avoid a proliferation of dimension six operators, whose self square contribution to $\mathcal O(1/\Lambda^4)$ results is not restricted by the need to interfere with the SM. Operators not on this list either have the wrong particle content, violate either the flavor or CP assumptions, or -- for dimension eight operators -- have the incorrect fermion helicity structure to interfere with the SM. We have used the naming convention from Ref.~\cite{Murphy:2020rsh} for all cases except  $Q^2L^2D^2$, where we place the derivatives in slightly different positions (as done in Ref.~\cite{Kim:2022amu}). For non-contact operators, we will use the definitions and naming convention from Refs.~\cite{Kim:2022amu}, rather than Ref.~\cite{Murphy:2020rsh}. The explicit form of these operators can be found in Appendix~\ref{app:opdefn}. }
\label{fig:theoperators}
\end{table}

\section{Amplitudes for $pp \to W^\pm(\ell^\pm) \gamma$}\label{sec:theamps}

In this section we provide the helicity amplitudes for $pp \to W^\pm(\ell^\pm) \gamma$ for the topologies shown in Fig.~\ref{fig:resonant}, \ref{fig:nonresonant}. We group the terms by the helicity of the final state photon and whether or not the diagram is resonant. Resonant diagrams follow a simple replacement procedure when swapping $+ \leftrightarrow -$ photon helicity, while off-resonant diagrams do not, due to the fact that the neutrino doesn't radiate. We take all fermions to be massless, using the `MCFM' convention~\cite{Bern:1996ka, Campbell:1999ah, Campbell:2011bn,Campbell:2015qma,Campbell:2019dru} for spinor helicity variables, and work with all momenta taken to be outgoing. 

We first consider $W^-$ production from quark ($p_2$) antiquark ($p_1$) including the $W^-$ decay to an electron ($p_3$) and anti-neutrino ($p_4$) plus a positive helicity photon ($p_5$). We will comment on how to convert to $W^+$ and negative helicity photons shortly. For brevity we will refer to the momenta by this numbering scheme, dropping the `$p$' in the following. 

First the SM-type diagrams, i.) and iii.). These contains the SM, plus all SMEFT input parameter effects and three-particle vertices with SM-like structure: 
\begin{align}
\mathcal A_{\rm SM-like}(1^-,2^+,3^-, 4^+,5^+)_{res} &=  -i\,2 \sqrt{2}\, \frac{\hat{p}_{34}}{s_{34}}\frac{\langle 13\rangle^2}{\langle 15\rangle}\frac{[34]}{ \langle 25\rangle}\Big(Q_d - \frac{\hat{p}_{12}}{s_{12}} (Q_d-Q_u)\, s_{25} \Big) \\
\mathcal A_{\rm SM-like}(1^-,2^+,3^-, 4^+,5^+)_{non. res} &= -i\,2 \sqrt{2}\, \frac{\langle 13\rangle^2}{\langle 15\rangle}\frac{ \hat p_{12}}{s_{12}} \frac{(Q_d-Q_u)[24]}{\langle 35\rangle} 
\end{align}
where 
\begin{align}
\hat p_{ij} = \frac{s_{ij}}{s_{ij} - m^2_W + i \Gamma_W m_W}. \nonumber
\end{align}
The sum of the two contributions can be massaged into the form found in MCFM~\cite{Dixon:1998py, Campbell:1999ah, Campbell:2011bn, Campbell:2021mlr}. Note that while SMEFT effects will enter into the coupling that accompany this amplitude, they will not introduce any new kinematics (and the same holds for all other amplitudes shown in this section).

SMEFT $VVV$ couplings include two structures beyond what's found in the SM. We'll refer to these as `$C_W$' type, and `$C_{HWB}$' type, naming them after the leading SMEFT operators which contribute. Sewing these $VVV$ vertices into the $\hat s$ channel piece of diagram set i.) we have
\begin{align}
\mathcal A_{\rm c_{HWB}}(1^-,2^+,3^-, 4^+,5^+)_{res} &=i\,\sqrt 2\, \frac{\hat p_{12}}{s_{12}} \frac{\hat p_{34}}{s_{34}} \langle 13\rangle[25][45]  \\
\mathcal A_{\rm c_{W}}(1^-,2^+,3^-, 4^+,5^+)_{res} &=-i\, \sqrt 2\, \hat p_{12} \hat p_{34} \frac{[24][25][45]}{[12][34]}.
\end{align}
Up to overall normalization, these match the anomalous triple gauge amplitudes in Ref.~\cite{DeFlorian:2000sg}. At dimension eight, a third type of anomalous $VVV$ interaction can enter, stemming from operators in the class $H^4XD^2$ and encapsulated in the $\kappa^A_{IJ}(\phi)$ metric of Ref.~\cite{Helset:2020yio}. However, for $W^+W^-\gamma$ the $\kappa^A_{IJ}$ terms have the identical amplitude (kinematic) structure as the $C_{HWB}$ type term, so we don't need a new amplitude for them.

The final resonant diagram (type ii.)) comes from the dimension eight contact terms. Some contact terms, specifically those coming from $Q^2 X H^2 D$ operators, lead to $pp \to W^-(\ell^-\nu) \gamma$ amplitudes that are the same as $A_{\rm c_{HWB}}$ minus the $s$-channel propagator piece, so we will not repeat their form here. There are two new structures arising at dimension eight. Connecting these into contributions to $pp \to W^-(\ell^-\nu) \gamma$, we have:
\begin{align}
\label{eq:contactres}
\mathcal A^a_{\rm contact}(1^-,2^+,3^-, 4^+,5^+)_{res} & = i\, \frac{\hat p_{34}}{s_{34}} \frac{[52]}{\sqrt 2}\Big( \langle 3 1 \rangle [54](s_{25}-s_{15}) + [52]\langle 2 1 \rangle (\langle 3 14] - \langle 3 2 4]) \Big)  \\
\mathcal A^b_{\rm contact}(1^-,2^+,3^-, 4^+,5^+)_{res} & =  i\,\frac{\hat p_{34}}{s_{34}} \frac{[52]}{\sqrt 2}\Big( \langle 3 1 \rangle [54](s_{25}-s_{15}) + [52]\langle 2 1 \rangle (\langle 3 14] + \langle 3 2 4]) +  \nonumber  \\
& ~~~~~~~~~~~~~~~~~~~~~~~~~~~~~~~~~~~~~~~~~~~~~2 \langle 1 3 \rangle [42] \langle 2 1\rangle [51] \Big) . 
\end{align}
Here, $\langle 3 1 4]$ etc. are a shorthand,  $\langle 3 1 4] = \langle 3 1 \rangle [14]$.

For resonant diagrams, the negative helicity terms can be found by making the following replacements in the above:
\begin{align}
1 \leftrightarrow 2, 3 \leftrightarrow 4, \langle \rangle \leftrightarrow [],Q_u \leftrightarrow Q_d \nonumber
\end{align}

Moving on to the non-resonant terms, we have the set of terms in diagram iv.). The form of these amplitudes depends on the nature of the four fermion vertex, specifically whether or not there are any derivatives present. For dimension six contact terms and dimension eight terms of the form $Q^2 L^2H^2$, the amplitude for a positive helicity photon is:
\begin{align}
\mathcal A_{\psi^4}(1^-,2^+,3^-, 4^+,5^+)_{non. res} & = -i\,\frac{4}{\sqrt 2} \frac{\langle 13\rangle^2}{\langle 15\rangle}\Big( (Q_d - Q_u) \frac{[24]}{\langle 3 5 \rangle} - Q_d \frac{[43]}{\langle 2 5 \rangle} \Big)
\end{align}
where we have included the fermion electric charges in the amplitude rather than in the couplings.

At dimension eight, there are two operators with four fermions and two derivatives involving the final states we are interested in (and can interfere with the SM), entries $Q^2L^2D^2$ in Table~\ref{fig:theoperators}. Their contributions to Fig.~\ref{fig:nonresonant} diagram iv.) contain extra powers of momenta:
\begin{align}
\mathcal A_{D^2\psi^4,s}(1^-,2^+,3^-, 4^+,5^+)_{non. res} & = i\,\frac{8}{\sqrt 2} \frac{\langle 13\rangle^2}{\langle 15\rangle}\Big( (Q_d - Q_u) \frac{[24]}{\langle 3 5 \rangle}s_{12} - Q_d \frac{[43]}{\langle 2 5 \rangle}  (s_{12} + s_{15}) \Big) \nonumber \\
\mathcal A_{D^2\psi^4,t}(1^-,2^+,3^-, 4^+,5^+)_{non. res} & = i\,\frac{4}{\sqrt 2} \frac{\langle 13\rangle^2}{\langle 15\rangle}\Big( (Q_d - Q_u) \frac{[24]}{\langle 3 5 \rangle}(s_{13} + s_{15} + s_{24}) - \nonumber \\
&  \quad\quad\quad\quad\quad\quad\quad\quad Q_d \frac{[43]}{\langle 2 5 \rangle}  (s_{13} + s_{24} + s_{45})  \Big). \nonumber
\end{align}

Operators of the type $Q^2 L^2 D^2 $ also contain five-particle vertices -- diagram v.) -- where the photon comes directly from the covariant derivative rather than being radiated off one of the external lines. Grouping the fermion charge with the amplitude, these are:
\begin{align}
\mathcal A_{D^2\psi^4,s, 5pt }(1^-,2^+,3^-, 4^+,5^+)_{non. res} & = -i\,\frac{8}{\sqrt{2}} \frac{ \langle 13\rangle [42] }{ \langle 15 \rangle }\, Q_u\, [52]\langle 2 1 \rangle \nonumber \\
\mathcal A_{D^2\psi^4,t, 5pt }(1^-,2^+,3^-, 4^+,5^+)_{non. res} & = -i\,\frac{4}{\sqrt{2}} \frac{\langle 13\rangle [42]}{\langle 15 \rangle}\, (Q_u [53]\langle 3 1 \rangle + Q_d [54]\langle 41 \rangle). \nonumber 
\end{align}
This class of diagrams also gets a contribution from $Q^2L^2X$ operators, with amplitude:
\begin{align}
\mathcal A_{\psi^4X, 5pt }(1^-,2^+,3^-, 4^+,5^+)_{non. res} & =  -i\,\sqrt 2\, \langle 3 1 \rangle [52][54] \nonumber 
\end{align} 

The last class of diagrams (vi.)) contain an interaction similar to the contact terms in Eq.~\eqref{eq:contactres}, except now the contact portion of the amplitude is on the lepton side, with the $W^\pm$ propagator stitched to the initial quark current. 
\begin{align}
\mathcal A^c_{contact }(1^-,2^+,3^-, 4^+,5^+)_{non. res} & =  i\,\frac{\hat p_{12}}{s_{12}} \frac{1}{\sqrt 2 \langle 1 5\rangle}\Big(\langle 1 5 2] \langle 3 1 \rangle [54](s_{35}-s_{45}) + \nonumber \\
& \langle 31 \rangle [54](\langle 1 3 2] - \langle 1 4 2])(s_{15} + s_{25})  - \langle 3 5 4]\langle 1 5 2] \nonumber (\langle 1 3 2] - \langle 1 4 2]) \Big) \\
\mathcal A^d_{contact }(1^-,2^+,3^-, 4^+,5^+)_{non. res} & = i\,\frac{\hat p_{12}}{s_{12}} \frac{1}{\sqrt 2 \langle 1 5\rangle} \Big( \langle 1 3 \rangle [4 2] ([5 2 1\rangle (s_{45}-s_{35}) + \nonumber \\
& ([5 3 1\rangle - [5 4 1 \rangle)(s_{15} + s_{25})) + \langle 3 5 4] \langle 1 5 2]([5 3 1\rangle - [5 4 1 \rangle) \Big) \nonumber
\end{align} 
Operators of type $L^2XH^2D$ also lead to diagrams of this type and generate amplitudes with the same structure as $\mathcal A_{\psi^4X, 5pt }$, though with an additional propagator:
\begin{align}
\mathcal A_{DL2H2X }(1^-,2^+,3^-, 4^+,5^+)_{non. res} & = -i\,\sqrt 2\, \frac{\hat p_{12}}{s_{12}}[52][54]\langle 31 \rangle \nonumber
\end{align}

 To derive the amplitudes for a negative helicity photon, the amplitudes involving a five particle vertex follow the same simple replacement rule as the resonant terms. The other non-resonant terms do not follow the rule (though they do follow a more complicated rule involving the complex conjugate of process calculated using right handed fermions so they must be listed explicitly. They are presented in Appendix~\ref{app:negativehel}. Similarly, the amplitudes for $W^+$ production can be obtained from the above -- both resonant and non-resonant -- by a combination of complex conjugation and swapping indices: 
 \begin{align}
 \mathcal A_{W^+}(1^-,2^+,3^-,4^+,5^\lambda) = -(\mathcal A_{W^-}(1^-,2^+,4^-,3^+,5^{\lambda}))^*
 \end{align}

\section{Coupling factors in SMEFT}\label{sec:coupling}

With the helicity amplitudes known, the next step is to establish the coupling factors for each and, importantly, what order in the SMEFT expansion the couplings arise at, as this determines whether we keep `self-square' pieces at $\mathcal O(1/\Lambda^4)$.

The coupling factors are summarized in the table below. We've listed the amplitude from the previous section, the product of couplings that accompany the amplitude, and the lowest order in the SMEFT expansion where the operator appears. All factors of $i$ have been incorporated into the amplitudes already.
\begin{table}[h!]
\centering
\begin{tabular}{|c|c|c|} \hline
amplitude & coupling factors & lowest SMEFT order \\ \hline
$A_{SM-like}$ & $g_{Wq}\, g_{W\ell}\,  e$ & $\mathcal O(1)$ \\ \hline
$A_{\rm c_{HWB}}$  & $g_{Wq}\, g_{W\ell}\, g_{HWB}$  &  $\mathcal O(1/\Lambda^2)$ \\ \hline
$A_{\rm c_{HWB}}$  & $g_{Wq}\, g_{W\ell}\, \kappa_{HWB}$  &  $\mathcal O(1/\Lambda^4)$ \\ \hline
$A_{\rm c_{HWB}}$  & $ g_{W\ell}\, g_{DQ^2H^2X}$  &  $\mathcal O(1/\Lambda^4)$ \\ \hline
$A_{\rm c_{W}}$  & $g_{Wq}\, g_{W\ell}\, g_{W^3}$  &  $\mathcal O(1/\Lambda^2)$ \\ \hline
$A^{a,b}_{contact}$  & $ g_{W\ell}\, g_{DQ^2X^2}$  &  $\mathcal O(1/\Lambda^4)$ \\ \hline \hline
$A_{\psi^4}$  & $ g_{Q^2L^2}\, e$  &  $\mathcal O(1/\Lambda^2)$ \\ \hline
$A_{D^2\psi^4,(s,t)}$  & $ g_{D^2Q^2L^2,s}\, e, g_{D^2Q^2L^2,t}\, e $  &  $\mathcal O(1/\Lambda^4)$ \\ \hline
$A_{D^2\psi^4,5pt}$  & $ g_{D^2Q^2L^2,s}\, e, g_{D^2Q^2L^2,t}\, e $  &  $\mathcal O(1/\Lambda^4)$ \\ \hline
$A_{\psi^4X}$  & $ g_{Q^2L^2X} $  &  $\mathcal O(1/\Lambda^4)$ \\ \hline
$A_{\psi^4X}$ & $g_{Wq} g_{DL^2H^2X}$ & $\mathcal O(1/\Lambda^4)$ \\ \hline
$A^{c,d}_{DL^2X^2,non-res}$ & $g_{Wq} g_{DL^2X^2}$ & $\mathcal O(1/\Lambda^4)$ \\ \hline
\end{tabular}
\caption{Amplitudes, their corresponding coupling factors, and the order in the SMEFT expansion where the coupling factors first enter. The double line separates the resonant (above) and non-resonant  (below) categories. The difference between $A_{D^2\psi^4,5pt}$ and $A_{D^2\psi^4,(s,t)}$ is whether or not the photon comes from the same vertex as the four fermions.  The coupling for the five-particle vertex contains a factor of $e$, but we have separated it so that $A_{D^2\psi^4,5pt}$ and $A_{D^2\psi^4,(s,t)}$ have the same coupling dependence.}
\label{tab:couplingfactors}
\end{table}
Note that $A_{\rm c_{HWB}}$ and $A_{\psi^4X}$ appear several times, as these structures are generated by multiple operators. For $A_{\rm c_{HWB}}$, two different $VVV$ metrics (using the geoSMEFT language) and, up to a propagator factor, contact terms from operators in the class $Q^2 X H^2D$ all contribute. Explicitly, the coupling-dressed term proportional to $A_{cHWB}$ is:
\begin{align}
\label{eq:chwbkin}
 g_{W\ell}\, \mathcal A_{\rm c_{HWB}}(1^-,2^+,3^-, 4^+,5^+)_{res}\Big( g_{Wq}\, (g_{HWB} - \kappa_{HWB}) - \frac{s_{12}}{\hat p_{12}}\,g_{DQ^2H^2X} \Big).
\end{align}
In later sections, we'll refer to this combination as `having $C_{HWB}$ type kinematics'. An analogous expression can be written for $A_{\psi^4X}$.

Each of the coupling factors in Table~\ref{tab:couplingfactors} can be expanded in terms of Wilson coefficients. To be more transparent, we'll list the expansions for the individual coupling components rather than the product. Before proceeding, we need to pick an EW input scheme. As we have $W^\pm$ propagators all over the place, and following the advice of Ref.~\cite{Brivio:2021yjb}, we will use the $\hat m_W$ scheme. The translation between observables and SM inputs and SMEFT coefficients that this leads to has been given explicitly in Ref.~\cite{Hays:2020scx}. In the following, we will also use the variable $x = v^2_T/\Lambda^2$, where $v_T$ is the true minimum of the Higgs potential including SMEFT effects. Its relation to $G_F$, extracted from muon decay, can be found in Ref.~\cite{Hays:2020scx}.
  
\begin{itemize}
\item $g_{Wq}$ and $g_{W\ell}$ are the $W^\pm$ couplings to left handed quarks and leptons respectively. Their SMEFT expansion can be found in Ref.~\cite{Kim:2022amu}, along with the expression for $\Gamma_W$.\footnote{Had we chosen the $\hat\alpha_{ew}$ scheme, $m_W$ becomes a derived quantity and has a SMEFT expansion. See Ref.~\cite{Kim:2022amu}.}
\item $g_{HWB}$ is a $VVV$ vertex with anomalous kinematics. As suggested by the name, it is affected by $C^{(6)}_{HWB}$, though the dependence is a bit subtle as $C^{(6)}_{HWB}$  (and its higher dimensional analogs) is also involved in determining the translation between the vector boson gauge and mass eigenstates. To keep things straight, we can work with the geoSMEFT metrics first, then expand. Done this way, we find:
\begin{align}
g_{HWB} = \bar e \Big((\sqrt{g^{11}})^2 \Big( \frac{g_2}{g_1}\langle g_{34}\rangle + \langle g_{33} \rangle \Big) -1 \Big),
\end{align}

When expanded to $\mathcal O(1/\Lambda^4)$, this yields:
\begin{align}
g_{HWB} &= x\, C^{(6)}_{HWB} \hat e\,  \frac{c_{\hat \theta}}{s_{\hat \theta}} + \frac{x^2}{4}\, \hat e\,\Big(\frac{c_{\hat \theta}}{s_{\hat \theta}} (C^{(6)}_{HWB}(4\, C^{(6)}_{HB} + C^{(6)}_{HD} + 4 C^{(6)}_{HW} - 2\sqrt 2\, \delta G^{(6)}_{F}) + \nonumber \\
& ~~~~~~~~~~~~~~~~~~~~~~~~~~~~~~~~~~~~~~~~~~~~~~~~~~~~~~~~~~~~~~~~~~~~~ 2\, C^{(8)}_{HWB} \Big) - 4\, C^{(8),2}_{HWB} \Big). \nonumber
\end{align}
Here we have used the convention~\cite{Helset:2020yio, Hays:2020scx} where couplings without bars are SM Lagrangian parameters, while couplings with bars are combinations of Lagrangian parameters and gauge boson metric entries that are most readily compared with experiment. Hatted couplings/angles ($c_{\hat \theta} = \cos{\hat\theta}, s_{\hat \theta} = \sin{\hat\theta}$) are experimental inputs\footnote{Within the $\hat m_W$ input scheme, $\hat v = 1/(2^{1/4}\hat G_F), \hat \theta = \text{sin}^{-1}(\sqrt{1- \hat m^2_W/\hat m^2_Z})\,, \hat e = 2\times 2^{1/4}\, \hat m_W \sqrt{\hat G_F} \hat s_\theta$. We use $\hat G_F = 1.1663787\times 10^{-5}\, \text{GeV}^{-2}, \hat m_Z = 91.1976\, \text{GeV},$ and $\hat m_W = 80.387\, \text{GeV}$~\cite{Corbett:2021eux}.}.

\item $g_{W^3}$ is the second type of anomalous $VVV$ interaction. It can also be expressed in geoSMEFT metric, all-orders form as:
\begin{align}
g_{W^3} = \frac{\bar e}{g_2}(\sqrt g^{11})^2\, \Big( 6\, \langle f_{123}\rangle + 2\, \langle f_{124} \rangle \frac{g_2}{g_1} \Big).
\end{align}

Once expanded to $\mathcal O(1/\Lambda^4)$, this becomes:
\begin{align}
g_{W^3} &= x\, \frac{6\, C^{(6)}_W\, s_{\hat \theta}}{\hat v^2} + x^2 \frac{3 s_{\hat \theta}(6\, C^{(6)}_{W}\, C^{(6)}_{HW} - 2\sqrt 2 C^{(6)}_{W}\,\delta G^{(6)}_F + C^{(8)}_W)}{\hat v^2}  \nonumber \\
&~~~~~~~~~~~~~~~~~~~~~~~~~~~~~~~~~~~~~~~ -\frac{3c_{\hat \theta}\, C^{(6)}_W(4\,C^{(6)}_{HWB} + C^{(6)}_{HD}c_{\hat \theta}/s_{\hat \theta})}{2\hat v^2} \nonumber
\end{align}

\item The final anomalous $VVV$ coupling comes from the $\kappa$ metric, which is non-zero only at dimension eight and above.
\begin{align}
\kappa_{HWB} = \frac{v^2_T}{2} (\sqrt g^{11})^2\, \bar e\, g_2\, \Big(\langle \kappa^3_{12}\rangle  + \langle \kappa^4_{12}\rangle \frac{g_2}{g_1}\Big)\Big|_{\mathcal O(x^2)} = -x^2\, \frac{\hat e^2\, (C^{(8)}_{HDHB}\,c_{\hat \theta} + C^{(8)}_{HDHW}\,s_{\hat \theta})}{8\, \hat s^2_\theta} \nonumber
 \end{align}

\item For the resonant contact terms, the operators in Table~\ref{fig:theoperators}  translate into coupling factors as:
\begin{align}
\label{eq:DQ2X2}
g_{DQ^2H^2X} &= -x^2\,\hat e\, \frac{(C^{(8),1}_{Q^2BH^2D} - i\, C^{(8),3}_{Q^2BH^2D})c_{\hat \theta} + (C^{(8),11}_{Q^2WH^2D} + i\, C^{(8),9}_{Q^2WH^2D})}{\sqrt 2\, s_{\hat \theta} \hat v^2} \nonumber \\
g^{(a)}_{DQ^2X^2} &= x^2 \frac{(C^{(8),2}_{Q^2WBD}-C^{(8),1}_{Q^2WBD})\,c_{\hat \theta} - i\, C^{(8),2}_{Q^2W^2D}\, s_{\hat \theta}}{\sqrt 2 \hat v^4} \\
g^{(b)}_{DQ^2X^2} &=  x^2 \frac{(C^{(8),2}_{Q^2WBD}+C^{(8),1}_{Q^2WBD})\,c_{\hat \theta} + i\, C^{(8),2}_{Q^2W^2D}\, s_{\hat \theta}}{\sqrt 2 \hat v^4} \nonumber 
\end{align}

\item Continuing to the couplings for interactions leading to non-resonant pieces:
\begin{align}
\label{eq:qcontact}
g_{Q^2L^2} =  x\, \frac{2\,C^{(6),3}_{Q^2L^2}}{\hat v^2} + x^2 \frac{(C^{(8),2}_{Q^2L^2H^2} - i\,C^{(8),5}_{Q^2L^2H^2}) }{\hat v^2} \quad, & \quad  g_{Q^2L^2X} =  -2\,x^2\, \frac{(C^{(8),5}_{L^2Q^2W}\, s_{\hat \theta} - i\, C^{(8),3}_{L^2Q^2B}\, c_{\hat \theta})}{\hat v^4} \nonumber \\
g_{D^2Q^2L^2,s} = -x^2\, \frac{2\,C^{(8),3,s}_{Q^2L^2}}{\hat v^4}\quad, &\quad g_{D^2Q^2L^2,t} = -x^2\, \frac{2\,C^{(8),3,t}_{Q^2L^2}}{\hat v^4} \nonumber 
\end{align}
The $g_{DL^2X^2}$ and $g_{DL^2H^2X}$ couplings are identical to Eq.~\eqref{eq:DQ2X2}, with quark fields replaced by leptons.\footnote{Explicitly, mapping onto Ref.~\cite{Murphy:2020rsh}, these are $C^{(8),1}_{L^2BH^2D}, C^{(8),3}_{L^2BH^2D}, C^{(8),11}_{L^2WH^2D}, C^{(8),9}_{L^2WH^2D}$ for $ g_{DL^2H^2X}$ and $C^{(8),1}_{L^2WBD}, C^{(8),2}_{L^2WBD}, C^{(8),2}_{L^2W^2D}$ for $g_{DL^2X^2}$.}

\end{itemize} 

Note the $ffW$ couplings and contact terms from $\psi^2X^2D, \psi^2XH^2D$, $\psi^4X$ and $\psi^4H^2$ ($\psi = Q,L$ or combinations of them) are complex, with imaginary pieces all entering at $\mathcal O(1/\Lambda^4)$.  In these cases, we show the sign for $W^-(\ell^-\nu)$ production; complex conjugating gives the expressions for $W^+(\ell^+\nu)$ production.

\section{Resonant diagrams: Polarizaton study}\label{sec:pol}

Before launching into the full $2 \to 3$ calculation, let us take a step back to the $2 \to 2$ process $\bar q q' \to W^\pm \gamma$ and explore how the various polarization subamplitudes within the SM and SMEFT vary with energy. We focus on the resonant pieces as part of the goal of this study is to understand how robustly high-energy diboson processes can be used to constrain the anomalous $C_W$ type $VVV$ coupling at the LHC. From this perspective, non-resonant pieces are a nuisance that we can hopefully mitigate with selection cuts (See Sec.~\ref{sec:results}). As a side benefit, this polarization study will help us build intuition about how various higher dimensional SMEFT operators enter, which we can port to other processes.

Restricted to resonant terms, there are only four type of SMEFT effects we need to consider: i.) those with SM kinematics, ii.) contributions where the $VVV$ vertex has $C_{HWB}$ kinematics (see Eq.~\eqref{eq:chwbkin}), iii.) contributions where the $VVV$ has $C_{W}$ kinematics ($C^{(6)}_W$ and its higher dimensional iterations), and iv.) contact terms from $Q^2X^2D$ class operators. Using the methods of Ref.~\cite{Hagiwara:1986vm}, it is straightforward to construct the $2 \to 2$ amplitudes for the different possible $W^\pm \gamma$ polarization combinations. We are most interested in the energy dependence of the various terms, and in particular how they behave at large $\hat s$. This dependence is shown below in Table~\ref{tab:polcombos} for the four different contribution types:
\begin{table}[h!]
\centering
\begin{tabular}{|c|c|c|c|c|} \hline
$\epsilon_\gamma \epsilon_W$ & SM-like & $C_{HWB}\, VVV$ & $C_{W}\, VVV$ & $Q^2X^2D$ terms \\ \hline
$++$ & $\frac{v^2}{\hat s}$ & $\frac{v^2}{\Lambda^2}$ & $\frac{\hat s}{\Lambda^2}$ & $\frac{\hat s^2}{\Lambda^4}$ \\ \hline
$+-$ & 1 & 0 & 0 & $\frac{\hat s^2}{\Lambda^4}$ \\ \hline
$+0$ & $\frac{v}{\sqrt{\hat s}}$ & $\frac{v \sqrt{\hat s}}{\Lambda^2}$  & $\frac{v \sqrt{\hat s}}{\Lambda^2}$ & $ \frac{v \hat s^{3 / 2}}{\Lambda^4}$ \\ \hline
\end{tabular}
\caption{Energy dependence of the amplitudes for different $q\bar q \to W\gamma$ polarization possibilities. Amplitudes with polarizations flipped $+ \leftrightarrow -$ are identical. We have removed all coupling factors, set $m_W \sim v$ and dropped all dependence on the scattering angle. The energy dependence shown is just the leading behavior in the limit of large $\hat s$, meaning each column may contain pieces suppressed by additional factors of $v^2/\Lambda^2$ or $v^2/\hat s$. The amplitudes in the first column are the purely SM pieces, with all SMEFT corrections coming with powers of $1/\Lambda^2$. }
\label{tab:polcombos}
\end{table}

There are several takeaways from Table~\ref{tab:polcombos}. First, the $C_{HWB}$ type contribution has weaker energy dependence than the other two SMEFT columns, so we will ignore it. Next, we see the well-known fact that the $C_{W}$ structure does not contribute to the dominant $+-$ SM polarization~\cite{Azatov:2016sqh}. As such, if we want to find an energy enhanced (meaning growing with $\hat s$) amplitude squared using $C^{(6)}_W$, we need to look to the squared terms, $|\mathcal A^{++}_{C_W}|^2, |\mathcal A^{+0}_{C_W}|^2 $. Of these, $|\mathcal A^{++}_{C_W}|^2$ has the strongest energy growth, $\sim \hat s^2/\Lambda^4$. If we forget the contact terms, all of which arise at dimension eight, this contribution should dominate the high energy regions of $\bar q q' \to W^\pm \gamma$. Or, flipping the logic, if we ignore dimension eight operators, the high energy regions of $\bar q q' \to W^\pm \gamma$ are the place to look for/constrain $C^{(6)}_{W}$ effects. Similar conclusions can be drawn for $W^+W^-$ and $W^\pm Z$, see Ref.~\cite{Degrande:2023iob}, and this behavior is evident in diboson SMEFT studies at dimension six~\cite{Butter:2016cvz, Green:2016trm, Zhang:2016zsp, Baglio:2017bfe,Ellis:2018gqa,Baglio:2018bkm,Baglio:2019uty,Bellan:2021dcy,Almeida:2021asy}.

However, once we admit dimension eight operators, and in particular the $Q^2X^2D$ contact type terms, other terms in the SMEFT contribution can have significant energy enhancement and cannot be ignored. In particular, the $Q^2X^2D$ terms contribute to the polarization combination, $+-$, where the SM is largest, thus the interference $\mathcal A_{SM}^{+-}\mathcal A^{*,+-}_{contact}$ has the same energy scaling, $\sim \hat s^2/\Lambda^4$ as the $C^{(6)}_{W}$ squared piece.  Thus, interpreting high energy regions of $\bar q q' \to W^\pm \gamma$ solely in terms of $C^{(6)}_W$ could be misleading. The fact that dimension eight pieces interfering with the SM can have the same energy growth as the largest (energy-wise) dimension six squared terms in other diboson processes has recently been pointed in Ref.~\cite{Degrande:2023iob}.

Finally, while the energy parametrics are the same in $|\mathcal A^{++}_{C_W}|^2$ and $\mathcal A_{SM}^{+-}\mathcal A^{*,+-}_{contact}$ are the same, they involve different polarizations. It would be interesting to pursue the degree to which these could be separated, either by their kinematic distributions~\cite{Bern:2011ie,Stirling:2012zt,Denner:2020bcz,Poncelet:2021jmj}  or via the use of `polarization taggers' ~\cite{De:2020iwq, Kim:2021gtv}.

Our study also clues us in to shortcuts for identifying which polarization combinations a particular operator will contribute to, using just the field content of the operator. The first shortcut is one we have already mentioned: transverse gauge bosons live primarily in field strengths, while longitudinal gauge bosons (which vanish in the limit $v \to 0$) reside primarily in $D_\mu H$ type terms. By primarily, we mean in the $\sqrt{\hat s} \gg m_W$ limit, where the price to convert from transverse to longitudinal (i.e. the amount of longitudinal $W^\pm$ in a field strength) is $\sim m_W/\sqrt{\hat s}$. 

To go further and understand $+/-$ polarization, let us revert the operators to a more group theoretical, rather than phenomenological, form. By this we mean we are only interested in the representations of the fields involved in the operators and not how indices are contracted. When looking at representations, field strengths are most conveniently expressed in $L$ and $R$ combinations $W_{L/R} = W_{\mu\nu} \pm i\, \tilde W_{\mu\nu}$, as these objects have simple Lorentz group representations, $W_L = (1,0) , W_R = (0,1)$ of $SU(2)_L \otimes SU(2)_R$. The Lorentz group representations tell us the particle's helicity, $+$ for $W_R$ and $-$ for $W_L$ (in our convention). In this group representation form, $C^{(6)}_W$ operators have the form $W^3_L$ or $W^3_R$, while $C^{(8)}_{Q^2X^2D}$ all involve $W_L W_R$ (or $B_L W_R$, etc.)\footnote{While Ref.~\cite{Murphy:2020rsh, Li:2020gnx} have bases in phenomenological form, lists of the group theoretical form used here can be found in Ref.~\cite{Lehman:2015coa, Henning:2015alf} based off of Hlibert series counting techniques.}. From this group structure, we expect $C^{(6)}_W$ will contribute strongest when both of the external gauge bosons have the same helicity/polarization while $Q^2X^2D$ will contribute primarily to opposite helicities/polarizations -- exactly the pattern seen in Table~\ref{tab:polcombos}. 

Analyzing $C^{(6)}_{HWB} \sim H^2 W_L B_L + h.c.$ from this angle, we see it also involves particles of the same helicity and contributes predominantly to $++$ and $--$. It's weaker energy scaling compared to $C^{(6)}_W$ comes from vev counting. By dimensional analysis, dimension six operators will contribute to a $2 \to 2$ amplitude $\sim v^2/\Lambda^2, v\sqrt{\hat s}/\Lambda^2$ or $\hat s/\Lambda^2$, and since we need to set both Higgses in $C^{(6)}_{HWB}$ to vevs to make a vertex with no Higgses, this exhausts the mass/energy powers in the numerator forcing us into the $v^2/\Lambda^2$ category. Operators without Higgses, such as $C^{(6)}_W$ or $C^{(8)}_{Q^2X^2D}$, can contribute to vertices relevant for $\bar q q' \to W^\pm \gamma$  without having to set any vevs, and therefore will have stronger energy growth.

\section{Results for $pp \to W^\mp(\ell^\mp\nu) \gamma$ }\label{sec:results}

In this section we present results for the full proton-level calculation of  $pp \to W^-(\ell^-\nu) \gamma$, $W^+(\ell^+\bar\nu) \gamma$. We dress each amplitude with the respective couplings, then square and sum over both photon helicities, making sure to retain pieces only out to $\mathcal O(1/\Lambda^4) = \mathcal O(x^2)$. By this we mean we retain only the interference terms for the amplitudes in Table~\ref{tab:couplingfactors} that first appear at $\mathcal O(1/\Lambda^4)$, but we retain both the interference and squared terms for amplitudes which begin at $\mathcal O(1/\Lambda^2)$. Interference between two different $\mathcal O(1/\Lambda^2)$ amplitudes is included in what we call the squared term.

The net amplitude squared is combined with phase space, initial state factors, and convolved with parton distribution functions to get the total integrand. To manipulate the spinor helicity products and convert them into Lorentz invariants, we use the packages SpinorHelicity4D~\cite{AccettulliHuber:2023ldr} and FeynCalc~\cite{Hahn:2000kx,Hahn:1998yk}. The integrands -- corresponding to various terms in the amplitude squared -- are then integrated using the method described in Ref.~\cite{Corbett:2023yhk}.  All numerical integration is handled by GSL~\cite{gough2009gnu} library and routines. We use the NNPDF3.0 parton distributions~\cite{Hartland_2013,Ball_2015}, interfaced to the code with LHAPDF~\cite{Buckley:2014ana}.  In the numerical results shown below, we use a factorization scale of $\mu_F = m_W$, $\alpha_s = 0.118$ and a collider center of mass energy of 13 TeV. In addition to determining the total cross section (as a function of the final state cuts -- which are necessary even in the SM to avoid a singularity associated with collinear photons), we also generate kinematic distributions using the reweighting technique from Ref.~\cite{Corbett:2023yhk}. This is straight forward for $pp \to W^\pm(\ell^\pm\nu)\gamma$ as only one fermion helicity combination contributes.

We make the following simplifications to shorten our results and focus on the physics we are most interested in. First, writing $\Gamma_W$ as a SMEFT expansion and expanding, we can generate new terms in the amplitude squared proportional to $\delta \Gamma_W$. In keeping with a consistent expansion, these effects can only accompany amplitudes whose coupling factors are $\mathcal O(x)$ or lower. Reference~\cite{Corbett:2023yhk} explored these terms in the context of $VH$ production and found them to be small\footnote{Since the width involves the same $ffV$ couplings that enter into the vertex, there is no way to adjust the Wilson coefficients to affect $\delta \Gamma$ alone (meaning leaving $g_{Wq}$ or $g_{W\ell}$ unchanged).}, so we will ignore them in the numerical study here. We emphasize that the full SMEFT dependence is captured in Sec.~\ref{sec:theamps} and \ref{sec:coupling} in the amplitudes and coupling factors, should the reader wish to retain the $\delta \Gamma$ effects. Second, we see that several SMEFT couplings are complex. In a cross section calculation, the real part of a product of couplings multiplies the real part of the corresponding product of (coupling-stripped) amplitudes, while the imaginary part of a coupling product multiplies the imaginary part of the amplitude product. The imaginary part of the amplitude product is proportional either to $\Gamma_W$ or $\epsilon_{\mu\nu\rho\sigma}$. Both of these are suppressed compared to other terms; $\Gamma_W$ is suppressed as $\Gamma_W \ll m_W, \sqrt{\hat s}$ (for the range of collision energies we are interested in), while $\epsilon_{\mu\nu\rho\sigma}$ is zero if integrated over the entire phase space\footnote{As we will impose cuts and look differentially, the $\epsilon_{\mu\nu\rho\sigma}$ pieces are not exactly zero, however we find them to be very small compared to the real part of the amplitude products}. Finally, we take the CKM matrix to be diagonal, as Yukawa couplings are absent in the strict $U(3)^5$ limit. Keeping $U(3)^5$ for all higher dimensional terms while allowing the SM ($d=4$) Yukawa interactions reinstates the CKM matrix for all (SM and SMEFT) $\bar q q'$ vertices. This would lead to effects of order $\mathcal O(V^2_{us})$ times the ratio of $\bar u s$ to $\bar u d$ parton luminosities, roughly $3\%$~\cite{Corbett:2023yhk}.

We impose the following parton level cuts, implemented  to avoid regions of phase space where the photon and lepton become collinear and to focus on the $\hat s \gg v$ regime where energy-enhanced SMEFT effects will be more pronounced.
\begin{align}
p_{T,\ell} \ge 10\, \text{GeV},\, &\, |\eta_\ell| \le 2.5 \nonumber \\
p_{T,\gamma} \ge 200\, \text{GeV},\, & \, |\eta_\gamma| \le 2.5  \\
\Delta R_{\ell,\gamma}&\ge 0.4 \nonumber
\label{eq:cuts}
\end{align}
The cross section for each of the contributions is shown below in Table~\ref{tab:CC}. The SM term is the first row, followed by the $\mathcal O(1/\Lambda^2)$ resonant pieces and their self-squares. Below those are five $\mathcal O(1/\Lambda^4)$ resonant pieces. All of the terms below the double line are the non-resonant contributions, grouped as i.) $\mathcal O(1/\Lambda^2)$ piece, ii.) the self squared $\mathcal O(1/\Lambda^4)$ piece, then iii.) all non-resonant $\mathcal O(1/\Lambda^4)$ interference pieces -- both SM$\times$ dimension eight and the interference between two different dimension six.  

\begin{table}
\centering
\addtolength{\leftskip} {-2cm}
\addtolength{\rightskip}{-2cm}
\renewcommand{\arraystretch}{1.4}
\begin{tabular}{|r|c|c|c|c|c|}
\hline
&\multicolumn{4}{c|}{partons}\\
\hline
($\mathcal L_{\rm eff}$ dependence)/$N_c$&$\bar u d$&$\bar c s$&$\bar d u$&$\bar d c$\\
\hline
\hline
$|g_{Wq}|^2\,|g_{W\ell}|^2\, e^2$ & 0.74 & 0.08 &  1.51 & 0.07 \\ 
\hline
\hline
$|g_{Wq}|^2\,|g_{W\ell}|^2\, e\, g_{HWB}$ & -0.72 & -0.86 & -1.16 & -1.15 \\
$|g_{Wq}|^2\,|g_{W\ell}|^2\, e\, g_{W^3}\, \hat v^2$ & 0.16 & 0.18 & 0.22 & 0.21 \\
\hline
$|g_{Wq}|^2\,|g_{W\ell}|^2\,  g^2_{HWB}$ & 17.65 & 16.82 & 17.95 & 16.2\\
$|g_{Wq}|^2\,|g_{W\ell}|^2\, g^2_{W^3}\, \hat v^4$ & 33.59 & 23.46 & 42.05 & 20.8 \\
\hline
$|g_{Wq}|^2\,|g_{W\ell}|^2\, g_{W3}\, g_{HWB}\, \hat v^2$ & -8.52 & -8.25 & -8.7  & -8.0 \\
$|g_{Wq}|^2\,|g_{W\ell}|^2\, e\, \kappa_{HWB}\, \hat v^4$ & 0.49 & 0.61 & 0.79  & 0.82 \\
$|g_{W\ell}|^2\, e\, {\rm Re}(g^*_{Wq} \, g_{DQ^2H^2X})\, \hat v^2 $ & 2.86 & 3.30 &  7.51 & 5.21 \\
$|g_{W\ell}|^2\, e\, {\rm Re}(g^*_{Wq}\, g^a_{DQ^2X^2})\, \hat v^4 $ & 40.10 & 26.48 & 43.5 & 19.4\\ 
$|g_{W\ell}|^2\, e\, {\rm Re}(g^*_{Wq}\, g^b_{DQ^2X^2})\, \hat v^4 $ & 36.79 & 26.54 & 39.1 & 20.0 \\ 
\hline
\hline
$e^2\, {\rm Re}(g^*_{Wq}\, g^*_{W\ell}\, g_{Q^2L^2})\, \hat v^2 $ & 0.93 & 0.68 & 3.55 & 1.92\\  
\hline
$e^2\, |g_{Q^2L^2}|^2\, \hat v^4 $  & 13.02 & 6.76 & 58.7 & 15.5 \\  
\hline
$e\, {\rm Re}(g_{Wq}\, g_{W\ell})\, g_{Q^2L^2}\,g_{HWB}\, \hat v^2$ & -0.29 & -0.23 & -0.32 & -0.22 \\  
$e\, {\rm Re}(g_{Wq}\, g_{W\ell})\, g_{Q^2L^2}\, g_{W^3}\, \hat v^4$ & 1.28 & 0.79 & 1.62 & 0.63\\  
$e^2\, {\rm Re}(g_{Wq}\, g_{W\ell}\, g_{D^2Q^2L^2,s}) \hat v^4$ & 12.76 & 8.13 & 126.2 & 41.1 \\ 
$e^2\, {\rm Re}(g_{Wq}\, g_{W\ell}\, g_{D^2Q^2L^2,t}) \hat v^4$ & -8.66 & -5.45 & -85.8 & -27.4\\ 
$e^2\, {\rm Re}(g_{Wq}\, g_{W\ell}\, g_{Q^2L^2X})\, \hat v^4$ & -0.03 & -0.02 & 0.009 & -0.005 \\ 
$e\, {\rm Re}(g_{Wq}\, g_{W\ell}\, g_{D^2Q^2L^2,s})\, \hat v^4$ & 5.58 & 2.76 & 7.58 & 1.92 \\ 
$e\, {\rm Re}(g_{Wq}\, g_{W\ell}\, g_{D^2Q^2L^2,t})\, \hat v^4$ & -1.44 & -0.58 & -1.94 & -0.31\\ 
$e\, |g_{Wq}|^2\, {\rm Re}(g_{W\ell}\, g^ c_{DL^2X^2})\, \hat v^4$ & 0.05 & 0.03 & 0.06 & 0.02 \\ 
$e\, |g_{Wq}|^2\, {\rm Re}(g_{W\ell}\, g^ d_{DL^2X^2})\, \hat v^4$ &-0.32 & -0.18 & -0.44 & -0.22 \\ 
$e\, |g_{Wq}|^2\, {\rm Re}(g_{W\ell}\, g_{DL^2H^2X})\, \hat v^4$ & -0.002 & -0.002  & -0.001& -0.001 \\ 
\hline
\end{tabular}
\caption{In the first column we show the coupling dependence for the different SMEFT contributions to $pp \to W^\mp \gamma \to \ell^\mp \nu \gamma$. All numbers are in ${\rm pb}$ and are shown relative to the `SM-like' term, which sits in the first row (assuming a single lepton flavor in the final state). Several entries in the first column include powers $\hat v$, which have been included to offset powers of $1/\hat v$ in the couplings. See the text for details on cuts, parton distribution function choice, and scale choices.}\label{tab:CC}
\end{table}

We follow the same presentation scheme as in Ref.~\cite{Corbett:2023yhk}, meaning all the numbers in the second row and below should be multiplied by the first row entry to get the true value. Said another way, the numbers show the cross section relative to the SM structure. As an example, the $\bar u d \to \ell^- \bar\nu$ cross section proportional to $|g_{Wq}|^2\,|g_{W\ell}|^2\,  g^2_{HWB}$ is:
\begin{align}
 |g_{Wq}|^2\,|g_{W\ell}|^2\,  g^2_{HWB}\times (0.74\, \text{pb})\times (17.65)
\end{align}
The cross sections are for an individual lepton, so to convert these numbers into $pp\to \ell^\mp \nu$ we need to sum the different partonic contributions and multiply by the number of lepton flavors considered.
Note that we have factored out powers of $\hat v$ in several of the rows to make the numbers more similar. These factors are often, but not always, set by the $\Lambda$ order of the couplings involved. More accurately, they are determined to offset powers of $1/\hat v$ in the couplings in Sec.~\ref{sec:coupling}.

Inspecting Table~\ref{tab:CC}, we see that several SMEFT amplitudes are enhanced relative to the SM. Among the resonant pieces, the enhanced terms include those with couplings $g^2_{W^3}$, $g^a_{DQ^2X^2}$ and $g^b_{DQ^2X^2}$, all of which is to be expected given the arguments in Sec.~\ref{sec:pol} and the fact that the cuts applied put us in a regime where $\hat s \gg v$. However, several of non-resonant terms are also large compared to the SM, notably the four fermion terms at dimension six ( coupling factor $|g_{Q^2L^2}|^2$) and the four fermion, two derivative terms at dimension eight (couplings $g_{D^2Q^2L^2,s}, g_{D^2Q^2L^2,t}$) (recall that the difference between the two occurrences of terms with coupling $g_{D^2Q^2L^2,s}, g_{D^2Q^2L^2,t}$ is whether or not the photon comes from the same vertex as the four fermions -- topologies of type iv.) vs. v.) in Fig.~\ref{fig:nonresonant}. If we assume all Wilson coefficients have roughly the same size, the fact that multiple SMEFT contributions are enhanced relative to the SM makes the task of disentangling any observed differences between high energy $pp \to W^\pm (\ell^\pm \nu) \gamma$ data and SM much trickier. 

To try to isolate different contributions we can impose cuts that focus on/off of the resonant $W^\pm$ region. As the neutrino is not observed, we cannot cut on an invariant mass so the best we can do is to place cuts on the transverse mass $m_{T, \ell\nu}$. To focus on the resonant region, we impose a cut $|m_{T,\ell \nu} - m_W| \le 20\, \text{GeV}$ in addition to the cuts in Eq.~\eqref{eq:cuts}. The results, presented in the same format as Table~\ref{tab:CC}, are shown below in Table~\ref{tab:CCcuts}.

\begin{table}
\centering
\renewcommand{\arraystretch}{1.4}
\begin{tabular}{|r|c|c|c|c|c|}
\hline
&\multicolumn{4}{c|}{partons}\\
\hline
($\mathcal L_{\rm eff}$ dependence)/$N_c$&$\bar u d$&$\bar c s$&$\bar d u$&$\bar d c$\\
\hline
\hline
$|g_{Wq}|^2\,|g_{W\ell}|^2\, e^2$ & 0.35  & 0.04 & 0.65 &  0.03 \\ 
\hline
\hline
$|g_{Wq}|^2\,|g_{W\ell}|^2\, e\, g_{HWB}$ & -0.19 & -0.28  & -0.44 & -0.48\\
$|g_{Wq}|^2\,|g_{W\ell}|^2\, e\, g_{W^3}\, \hat v^2$ & -1.25 & -1.17 & -1.35 & -1.16\\
\hline
$|g_{Wq}|^2\,|g_{W\ell}|^2\,  g^2_{HWB}$ & 17.45 & 16.48 & 19.3 & 16.8 \\
$|g_{Wq}|^2\,|g_{W\ell}|^2\, g^2_{W^3}\, \hat v^4$ & 33.17 & 23.31 & 45.0 & 22.2 \\
\hline
$|g_{Wq}|^2\,|g_{W\ell}|^2\, g_{W^3}\, g_{HWB}\, \hat v^2$ & -9.06 & -8.72 & -10.05  & -8.91 \\
$|g_{Wq}|^2\,|g_{W\ell}|^2\, e\, \kappa_{HWB}\, \hat v^4$ & 0.26 & 0.32  & 0.44 & 0.48 \\
$|g_{W\ell}|^2\, e\, {\rm Re}(g^*_{Wq} \, g_{DQ^2H^2X})\, \hat v^2 $ & 1.01 & 1.46  & 3.41 & 2.67 \\
$|g_{W\ell}|^2\, e\, {\rm Re}(g^*_{Wq}\, g^a_{DQ^2X^2})\, \hat v^4 $ & 50.33 & 33.49  & 60.0 & 26.7 \\ 
$|g_{W\ell}|^2\, e\, {\rm Re}(g^*_{Wq}\, g^b_{DQ^2X^2})\, \hat v^4 $ &  23.34 & 17.22  & 24.4 & 13.0 \\ 
\hline
\hline
$e^2\, {\rm Re}(g^*_{Wq}\, g^*_{W\ell}\, g_{Q^2L^2})\, \hat v^2 $ & 0.07  & 0.07  & 0.25  & 0.26 \\  
\hline
$e^2\, |g_{Q^2L^2}|^2\, \hat v^4 $  & 0.34 & 0.26 & 1.17 & 0.79 \\  
\hline
$e\, {\rm Re}(g_{Wq}\, g_{W\ell})\, g_{Q^2L^2}\,g_{HWB}\, \hat v^2$ & -0.22 & -0.20 & -0.26  & -0.21 \\  
$e\, {\rm Re}(g_{Wq}\, g_{W\ell})\, g_{Q^2L^2}\, g_{W^3}\, \hat v^4$ & 0.17 & 0.15 & 0.22 & 0.15 \\  
$e^2\, {\rm Re}(g_{Wq}\, g_{W\ell}\, g_{D^2Q^2L^2,s})\, \hat v^4$ & 0.10 & 0.18 & 3.11 & 2.66 \\ 
$e^2\, {\rm Re}(g_{Wq}\, g_{W\ell}\, g_{D^2Q^2L^2,t})\, \hat v^4$ & -0.11 & -0.15 & -2.28 & -1.8 \\ 
$e^2\, {\rm Re}(g_{Wq}\, g_{W\ell}\, g_{Q^2L^2X})\, \hat v^4$ & 0.01 & 0.01 & 0.03 & 0.01 \\ 
$e\, {\rm Re}(g_{Wq}\, g_{W\ell}\, g_{D^2Q^2L^2,s})\, \hat v^4$ & 0.41 & 0.29 & 0.61 & 0.29\\ 
$e\, {\rm Re}(g_{Wq}\, g_{W\ell}\, g_{D^2Q^2L^2,t})\, \hat v^4$ & -0.08 & -0.04 & -0.10 & -0.03 \\ 
$e\, |g_{Wq}|^2\, {\rm Re}(g_{W\ell}\, g^ c_{DL^2X^2})\, \hat v^4$ & -0.01  & -0.004  & -0.01 & -0.005 \\ 
$e\, |g_{Wq}|^2\, {\rm Re}(g_{W\ell}\, g^ d_{DL^2X^2})\, \hat v^4$ & -0.025 & -0.018  & -0.04 & -0.02 \\ 
$e\, |g_{Wq}|^2\, {\rm Re}(g_{W\ell}\, g_{DL^2H^2X})\, \hat v^4$ &  $3\times10^{-4}$ & $10^{-4}$   & 0.001 & $4\times 10^{-4}$ \\ 
\hline
\end{tabular}
\caption{Identical setup as Table~\ref{tab:CC} except we have imposed an additional cut on the transverse mass of the $\ell^\pm\nu$ system, $|m_{T,\ell\nu}-m_W| \le 20\, \text{GeV}$. All numbers in the second and lower rows are shown relative to the `SM-like' term, which sits in the first row (given in ${\rm pb}$ assuming a single lepton flavor in the final state).}\label{tab:CCcuts}
\end{table}

Comparing Table~\ref{tab:CCcuts} to Table~\ref{tab:CC}, we see that the SM is reduced by about $50\%$. The values for the resonant SMEFT subprocess (relative to the new SM value) are similar to the values without the $m_T$ cut. In some cases, the relative value is larger. This can be explained by the fact that the SM-like term contains both resonant and non-resonant pieces, and is therefore more affected by the $m_T$ cut, than the purely resonant SMEFT terms. The non-resonant SMEFT terms are much smaller with the $m_T$ cut, with values roughly two orders of magnitude smaller than in Table~\ref{tab:CC}.

To further illustrate the impact of the $m_T$ cut, as well as to understand how various SMEFT contributions affect $pp \to W^\mp(\ell^\mp \nu) + \gamma$ in more detail, we turn to differential distributions. To form differential distributions, we produce a sample of $pp \to W^-(\ell^- \nu) + \gamma$ events, where each event (set of initial/final four vectors) is accompanied by an array of 22 weights, one for each of the rows in Table~\ref{tab:CC}. These weights are generated using the method described in Ref.~\cite{Corbett:2023yhk} (itself inspired by  the reweight~\cite{Artoisenet:2010cn} approach in MadGraph\cite{Alwall:2014hca}), and boils down to evaluating generated phase spaces point for each of the 22 matrix elements squared. Combining the numerical weight for each matrix element squared with the corresponding coupling factors and summing over all 22 contributions, we get a weight for each event as a function of the Wilson coefficients (as well as $x = \hat v/\Lambda$, and inputs such as $\hat e, s_{\hat \theta}$, etc.). Feeding that total weight into weighted histograms for any (parton level) kinematic variable, we can analyze how the distribution shape changes as we vary the Wilson coefficients.

\begin{figure*}
\addtolength{\leftskip} {-2cm}
\addtolength{\rightskip}{-2cm}
\centering
\includegraphics[width=0.5\textwidth]{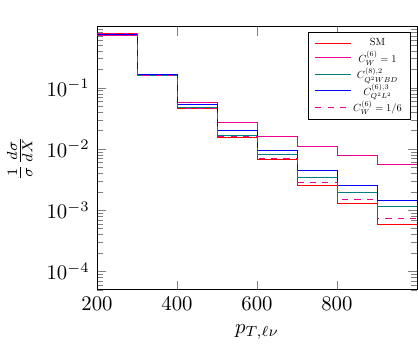}
\includegraphics[width=0.5\textwidth]{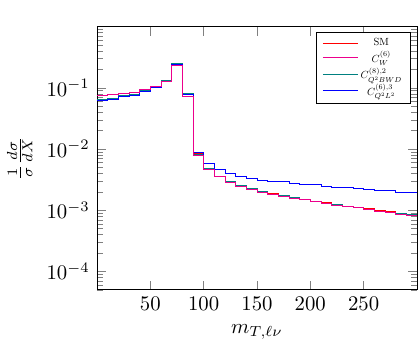}
\includegraphics[width=0.5\textwidth]{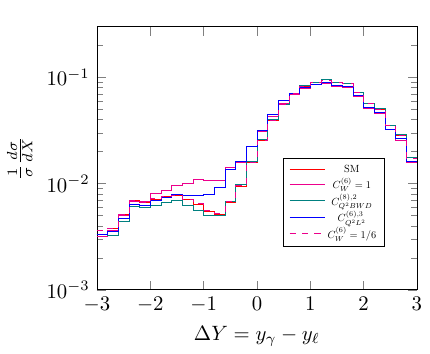}
\caption{Area normalized distributions of $p_{T,\ell\nu}$ (top panel) and $m_{T,\ell\nu}$ (middle panel) and $\Delta Y = y_\gamma - y_\ell$ (lower panel) in the SM (red solid line) and including three different Wilson coefficients. For the SMEFT plots, we set $\Lambda = 3\, \text{TeV}$ with the displayed coefficient set to $\pm 1$ (unless otherwise specified), and all other Wilson coefficients set to zero. $X$ on the vertical axis refers to the kinematic quantity on the horizontal axis.}
\label{fig:dist1}
\end{figure*}

In Fig.~\ref{fig:dist1} below we show three distributions generated with the basic cuts: $p_{T,\gamma}$, the transverse mass $m_{T,\ell\nu}$, and $y_\gamma - y_\ell$.  All distributions are area normalized in order to focus on shape differences. To study the SMEFT effects, we turn on one operator at a time. In the plot below, we show distributions for i.) $C^{(6)}_W$, the dimension six, triple-$W$ operator, ii.) $C^{(6),3}_{LQ}$ the dimension six, four fermion operator that includes a product of charged currents,  and iii.) $C^{(8),2}_{Q^2WBD}$ the dimension eight contact term with two field strengths (see Table~\ref{fig:theoperators}). We chose these three operators as they span the set of energy enhanced SMEFT contributions. 

Inspecting the $p_{T,\ell\nu}$ plot -- a proxy for $\sqrt{\hat s} $ -- we see all three operators lead to enhancements in the kinematic tails. For $C^{(8),2}_{Q^2WBD}$, the fact that the SMEFT is enhanced relative to the SM is sign dependent and actually requires a negative coefficient (with our operator convention), while the effect for dimension six terms is independent of the sign of $C^{(6)}_W, C^{(6),3}_{QL}$. Taking all coefficients equal to $\pm1$ with a scale $\Lambda = 3\, \text{TeV}$, the effect from $C^{(6)}_W$ is largest, though that is partially due to our choice of normalization. We have taken the operator to be $C^{(6)}_W\, \epsilon_{IJK} W_\mu^{I,\nu}\,W_\nu^{J\rho}W_\rho^{K\mu}$, which generates coupling factor $\sim 3!\, C^{(6)}_{W}$. Normalizing the operator to absorb the factorial from repeated fields, the effect of $C^{(6)}_W$ is much closer to the other operators shown (which are unaffected by the normalization change as they don't involve repeated fields). The distribution for $C^{(6)}_W = 1/6$ is shown in the dashed magenta line in the $p_{T,\ell\nu}$ plot; the fact that the normalization change is so dramatic is because the energy enhanced terms from $C^{(6)}_W$ come from its self square and not from interference with the SM. 

In the middle panel we show $m_{T,\ell\nu}$, the transverse mass of the lepton-neutrino system. We see that the $C^{(6)}_W, C^{(8),2}_{Q^2WBD}$ (area normalized) curves are indistinguishable from the SM, while the $C^{(6),3}_{Q^2L^2}$ curve is enhanced at large $m_{T, \ell\nu}$\footnote{There are some small differences between the $C^{(6)}_W$ and the SM $m_T$ distributions. $C^{(6)}_W$ does not interfere with the SM and therefore does not feel it's off-resonance contributions (topology iii.) in Fig.~\ref{fig:resonant}, while $C^{(8),1}_{DQ2X2}$ does. These differences are not visible with the plot zoomed out as shown. As such, we have only shown results for $C^{(6)}_W = 1$ }. This is expected, as the $C^{(6)}_W, C^{(8),2}_{Q^2WBD}$ SMEFT contributions are resonant, while $C^{(6),3}_{Q^2L^2}$ is not. Clearly, imposing the cut $|m_{T,\ell\nu} - m_W| < 20\, \text{GeV}$ will retain the former two SMEFT effects relative to the SM, while removing a larger amount of $C^{(6),3}_{Q^2L^2}$. To see the individual shapes of the contributions, rather than the stacked format see Appendix~\ref{app:mTshape}.

The last distribution in Fig.~\ref{fig:dist1} is the rapidity difference between the photon and the lepton $\Delta Y = y_\gamma - y_\ell$. We chose this distribution as it manifests aspects of the `radiation zero' in (tree-level) $W^\pm \gamma$ production\footnote{A more dramatic plot would be $\cos\theta^*$, where $\theta^*$ is the angle between the incoming quark and the outgoing photon. However, this is not viable at the LHC as we do not know which direction the initial quark came from.}, seen as the dip in the SM curve around $\Delta Y = -1$. The radiation zero is a natural place to look for new physics as the SM background is low~\cite{Capdevilla:2019zbx}. Honing in on this region, all of the operators we turned on predict deviations from the SM shape. Note that the deviations from the two resonant terms $C^{(6)}_W$ (either with coefficient 1 or $1/6$) and $C^{(8),2}_{Q^2WBD}$ have different shapes, which we can again trace back to the fact that they enter the $\mathcal O(1/\Lambda^4)$ cross section differently. The shape difference between $C^{(6)}_W$ and $C^{(8),2}_{Q^2WBD}$ means $\Delta Y$ could be use to disentangle any observed excess in $p_{T,\ell\nu}$. The non-resonant operator $C^{(6),3}_{QL}$ mimics $C^{(6)}_W$, however as we have illustrated in Table~\ref{tab:CCcuts} and the middle panel of Fig.~\ref{fig:dist1}, this contamination can be suppressed by an $m_{T,\ell\nu}$ cut (at least in the $\sqrt{\hat s} \gg v$ regime).

\section{Conclusions}\label{sec:conclude}

In this paper we have calculated the SMEFT helicity amplitudes for $\bar q q \to W^\pm(\ell^\pm \nu) + \gamma$ production up to $\mathcal O(1/\Lambda^4)$. This task is expedited by working with the geoSMEFT construction, where operators contributing to two and three-particle vertices have been worked out to all orders in $1/\Lambda$ and grouped into `metrics' dressing kinematic structures. Our calculation includes the decay of $W^\pm \to \ell^\pm \nu$ to facilitate comparison with experimental data and to study the effects non-resonant (meaning where the $\ell^\pm \nu$ do not originate from a $W^\pm$) SMEFT effects. We break down $\bar q q' \to W^\pm\gamma$ into different polarization combinations to determine which (resonant) SMEFT operators lead to cross sections with the strongest energy enhancement (in the limit $\sqrt{\hat s} \gg v$). The largest energy enhancement comes in the $++/--$ polarization channel from the self-square of amplitudes including the dimension-six triboson operator $C^{(6)}_W\, \epsilon_{IJK} W_\mu^{I,\nu}\,W_\nu^{J\rho}W_\rho^{K\mu}$ as well as in the $+-/-+$ channel from the interference between the SM and amplitudes including dimension eight $Q^2DX^2$ type operators. As both sources have the same energy dependence, $\mathcal O(\hat s^2/\Lambda^4)$, interpreting experimental data in terms of only one type of operator may be misleading. We then numerically explored the $\sqrt{\hat s} \gg v$ regime at the full $2 \to 3$ level. We find that several of the non-resonant SMEFT contributions are also energetically enhanced, however this `pollution' from non-resonant effects can be tempered by cuts on the transverse mass of the $\ell^\pm-\nu$ system. To supplement the fiducial cross section calculations and better illustrate the effects of different interesting SMEFT operators, we provide several differential distributions.

\section*{Acknowledgements}

We thank Tae Kim for collaboration during the early stages of this project, and Tyler Corbett for comments on the manuscript. The work of AM is partially supported by the National Science Foundation under Grant Number PHY-2112540.

\appendix 

\section{Definitions and naming conventions for non-contact operators}\label{app:opdefn}

For operators contained within geoSMEFT metrics, we use the naming conventions from Ref.~\cite{Helset:2020yio, Hays:2020scx}. We repeat them here for convenience. In the following, $n$ is a positive integer controlling the number of $(H^\dag H)$ factors present and $\tau^I$ are Pauli matrices. For operators with fermions $\psi$ labels the type of fermion, while $p,r$ label the fermion generation. We include the later for completeness and to match previous literature, as our flavor assumption implies the fermion interactions are universal among generations. \\

Bosonic operators with three field strengths:
\begin{align}
& C_W^{(6+2 n)}\, \epsilon_{IJK}\left(H^{\dagger} H\right)^n W_{\mu \nu}^I W^{\nu \rho, J} W_\rho{ }^{\mu, K} \\
& C_{W 2}^{(8+2 n)}\,\epsilon_{IJK}\left(H^{\dagger} H\right)^n\left(H^{\dagger} \tau^I H\right) W_{\mu \nu}^J W^{\nu \rho, K} B_\rho{ }^\mu
\end{align}

Bosonic operators with two field strengths ($g_{A B}$ metric):
\begin{align}
& C_{H B}^{(6+2 n)}\,\left(H^{\dagger} H\right)^{n+1} B^{\mu \nu} B_{\mu \nu} \\
& C_{H W}^{(6+2 n)}\,\left(H^{\dagger} H\right)^{n+1} W_I^{\mu \nu} W_{\mu \nu}^I \\
& C_{H W B}^{(6+2 n)}\,\left(H^{\dagger} H\right)^n\left(H^{\dagger} \tau^I H\right) W_I^{\mu \nu} B_{\mu \nu} \\
& C_{H W, 2}^{(8+2 n)}\,\left(H^{\dagger} H\right)^n\left(H^{\dagger} \tau^I H\right)\left(H^{\dagger} \tau^J H\right) W_I^{\mu \nu} W_{J, \mu \nu} 
\end{align}

Bosonic operators with a single field strength ($k_{I J}^A$ metric):
\begin{align}
& C_{H D H B}^{(8+2 n)}\.i\left(H^{\dagger} H\right)^{n+1}\left(D_\mu H\right)^{\dagger}\left(D_\nu H\right) B^{\mu \nu} \\
& C_{H D H W}^{(8+2 n)}\,i \delta_{I J}\left(H^{\dagger} H\right)^{n+1}\left(D_\mu H\right)^{\dagger} \tau^I\left(D_\nu H\right) W_J^{\mu \nu} \\
& C_{H D H W, 2}^{(8+2 n)}\, i \epsilon_{I J K}\left(H^{\dagger} H\right)^n\left(H^{\dagger} \tau^I H\right)\left(D_\mu H\right)^{\dagger} \tau^J\left(D_\nu H\right) W_K^{\mu \nu},
\end{align}

Bosonic operators without field strengths ($h_{IJ}$ metric):
\begin{align}
& C_{H D}^{(8+2 n)}\,\left(H^{\dagger} H\right)^{n+2}\left(D_\mu H\right)^{\dagger}\left(D^\mu H\right) \\
& C_{H, D 2}^{(8+2 n)}\,\left(H^{\dagger} H\right)^{n+1}\left(H^{\dagger} \tau_I H\right)\left(D_\mu H\right)^{\dagger} \tau^I\left(D^\mu H\right)
\end{align}

Fermion bilinears ($L_{J, A}^{\psi, p r}$ metric):
\begin{align}
& C_{p r}^{1,(6+2 n)}\,\left(H^{\dagger} H\right)^n H^{\dagger} \overleftrightarrow{i D^\mu} H \bar{\psi}_p \gamma_\mu \psi_r \\
& \underset{p r}{C_H^{3,(6+2 n)}}\,\left(H^{\dagger} H\right)^n H^{\dagger} i\,\overleftrightarrow{D_I^\mu} H \bar{\psi}_p \gamma_\mu \tau^I \psi_r \\
& C_{p r}^{2,(8+2 n)}\,\left(H^{\dagger} H\right)^n\left(H^{\dagger} \tau_I H\right) H^{\dagger} i\,\overleftrightarrow{D^\mu} H \bar{\psi}_p \gamma_\mu \tau^I \psi_r \\
& \underset{p r}{C_H^{\epsilon,(8+2 n)}}\,\epsilon^{I J K}\left(H^{\dagger} H\right)^n\left(H^{\dagger} \tau_K H\right) H^{\dagger} i\, \overleftrightarrow{D}_J^\mu H \bar{\psi}_p \gamma_\mu \tau_I \psi_r 
\end{align}
Lastly, $\delta G^{(6)}_F$ contains corrections to the Fermi constant coming from four-fermion contact operators and deviations in the $ffW$ vertex for muons or electrons:
\begin{align}
\label{eq:deltaGF}
 \delta G_F^{(6)} &= \frac{1}{\sqrt2} \left(C^{(6),3}_{\substack{Hl \\ee}}+ C^{(6),3}_{\substack{Hl \\ \mu \mu}} - \frac{1}{2}( C'_{\substack{ll \\ \mu ee \mu}}+ C'_{\substack{ll \\ e \mu \mu e}})\right).
\end{align} 

\section{Nonresonant Amplitudes for negative helicity photons} \label{app:negativehel}

For resonant diagrams, the negative helicity amplitudes for $q\bar q \to W^- (\ell^-\nu) + \gamma$ can be obtained from the positive helicity results shown in Sec.~\ref{sec:theamps} by replacing $1 \leftrightarrow 2, 3 \leftrightarrow 4, \langle \rangle \leftrightarrow [],Q_u \leftrightarrow Q_d \nonumber$. For the nonresonant diagrams, this replacement doesn't work (as the neutrino line does not emit a photon), so we present those amplitudes here.

\begin{align}
\mathcal A_{\rm SM-like}(1^-,2^+,3^-, 4^+,5^-)_{non. res} &= -i\,2 \sqrt{2}\, \frac{[ 24]^2}{[52]}\frac{ \hat p_{12}}{s_{12}} \frac{(Q_d-Q_u)\langle 14\rangle}{[53]} 
\end{align}

\begin{align}
\mathcal A_{\psi^4}(1^-,2^+,3^-, 4^+,5^-)_{non. res} & = -i\,\frac{4}{\sqrt 2} \frac{[24]^2}{[52]}\Big(  Q_u \frac{\langle 43 \rangle}{ [51]} - (Q_d - Q_u) \frac{\langle 41 \rangle}{[53]}\Big)
\end{align}

\begin{align}
\mathcal A_{D^2\psi^4,s}(1^-,2^+,3^-, 4^+,5^-)_{non. res} & = i\,\frac{8}{\sqrt 2} \frac{[24]^2}{[52]}\Big(  Q_u \frac{\langle 43 \rangle}{ [51]}(s_{12} + s_{25}) + (Q_d - Q_u) \frac{\langle 41 \rangle}{[53]} s_{12}\Big)  \nonumber \\
\mathcal A_{D^2\psi^4,t}(1^-,2^+,3^-, 4^+,5^-)_{non. res} & = i\,\frac{4}{\sqrt 2} \frac{[24]^2}{[52]}\Big(  Q_u \frac{\langle 43 \rangle}{ [51]}(s_{13} + s_{35} + s_{24}) + \nonumber \\
 &  \quad\quad\quad\quad\quad\quad\quad (Q_d - Q_u)\frac{\langle 41 \rangle}{[53]}(s_{13} + s_{15} + s_{24})\Big)   
\end{align}

\begin{align}
\mathcal A_{D^2\psi^4,s, 5pt }(1^-,2^+,3^-, 4^+,5^-)_{non. res} & = -i\,\frac{8}{\sqrt{2}} \frac{ \langle 13\rangle [42] }{ [5 2] }\, Q_d\, \langle 51 \rangle [12]   \\
\mathcal A_{D^2\psi^4,t, 5pt }(1^-,2^+,3^-, 4^+,5^-)_{non. res} & = -i\,\frac{4}{\sqrt{2}} \frac{\langle 13\rangle [42]}{[52]}\, (Q_u \langle 53\rangle [3 2] + (Q_d - Q_u)\langle 5 1 \rangle [1 2] \nonumber \\
&\quad\quad\quad\quad\quad\quad\quad\quad\quad\quad\quad\quad\quad + Q_d \langle 54\rangle [42]) 
\end{align}

\begin{align}
\mathcal A_{\psi^4X, 5pt }(1^-,2^+,3^-, 4^+,5^-)_{non. res} & =  -i\,\sqrt 2\, [25]\langle 1 5 \rangle \langle 3 5 \rangle
\end{align} 

\begin{align}
\mathcal A^c_{contact }(1^-,2^+,3^-, 4^+,5^-)_{non. res} & =  i\,\frac{\hat p_{12}}{s_{12}} \frac{1}{\sqrt 2 [52]}\Big(\langle 1 5 2] \langle 3 5 \rangle [24](s_{35}-s_{45}) +  \\
& \langle 35 \rangle [24](\langle 1 3 2] - \langle 1 4 2])(s_{15} + s_{25})  - \langle 3 5 4]\langle 5 1 2] \nonumber (\langle 1 3 2] - \langle 1 4 2]) \Big) \\
\mathcal A^d_{contact }(1^-,2^+,3^-, 4^+,5^-)_{non. res} & = i\,\frac{\hat p_{12}}{s_{12}} \frac{1}{\sqrt 2 [52]} \Big( \langle 1 3 \rangle [4 2] (\langle 5 1 2] (s_{45}-s_{35}) +  \\
& (\langle 5 3 2] - \langle 5 4 2])(s_{15} + s_{25})) + \langle 3 5 4] \langle 1 5 2](\langle 5 3 2] - \langle 5 4 2]) \Big) \nonumber
\end{align} 

\begin{align}
\mathcal A_{DL2H2X }(1^-,2^+,3^-, 4^+,5^-)_{non. res} & = -i\,\sqrt 2\, \frac{\hat p_{12}}{s_{12}}\langle 15 \rangle \langle 35 \rangle [24] 
\end{align}

\section{Shape analysis of $m_{T,\ell\nu}$}\label{app:mTshape}

\begin{figure*}[h!]
\centering
\includegraphics[width=0.75\textwidth]{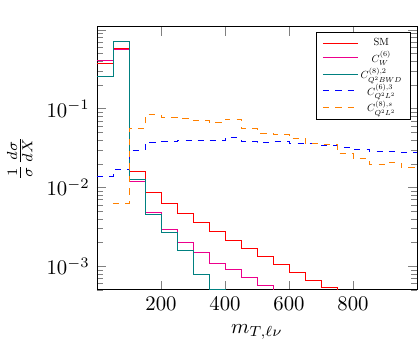}
\caption{Area normalized distributions of $m_T$ for several of the contributions in Table~\ref{tab:CC}, zoomed out to see a wider range. The SM-like curve (red) and the two resonant contributions, $\sim g^2_{W^3}$ (magenta) and $\sim g^a_{DQ^2X^2}$ (green)  fall precipitously after $m_T \sim m_W$ while the non-resonant contributions, $\sim |g_{Q^2L^2}|^2$ (dashed blue) and $\sim g_{D^2Q^2L^2,s}$ (orange) have the bulk of their support at $m_T \gg m_T$.}
\label{fig:mTshape}
\end{figure*}

To further understand the impact of the $|m_{T,\ell\nu} - m_W|$ cut, in Figure~\ref{fig:mTshape} we show the $m_T$ distributions from several of the individual contributions listed in Table~\ref{tab:CC}. The distributions for the individual matrix element contributions are superimposed and area normalized (rather than stacked on top of the SM as in Fig.~\ref{fig:dist1}) to emphasize their different shape. The curve for the SM-like matrix element is peaked at $m_T \sim m_W$ and falls steeply afterwards. The curves associated with the resonant SMEFT contributions follow the same pattern, falling even more steeply than the SM, while the non-resonant SMEFT terms have broad $m_T$ distributions. 
Truncating $m_T$ close to $m_W$ we can see that only a small fraction of the non-resonant SMEFT pieces are retained.

\bibliographystyle{JHEP}
\bibliography{ref.bib}

\end{document}